\documentclass[fleqn,usenatbib]{mnras}

\usepackage{newtxtext,newtxmath}

\usepackage[T1]{fontenc}

\DeclareRobustCommand{\VAN}[3]{#2}
\let\VANthebibliography\thebibliography
\def\thebibliography{\DeclareRobustCommand{\VAN}[3]{##3}\VANthebibliography}

\usepackage{graphicx}	
\usepackage{amsmath}	

\usepackage{longtable}

\usepackage{color}
\usepackage{booktabs}
\usepackage{siunitx}
\usepackage{multicol}
\usepackage{multirow}
\usepackage{float}
\usepackage{pdflscape}	
\usepackage{tabulary,booktabs}

\usepackage[dvipsnames]{xcolor}

\title[AGN torus disappearance]{Observational hints on the torus obscuring gas behaviour through X-rays with \textit{NuSTAR} data}

\author[Osorio-Clavijo et. al. ]{
N. Osorio-Clavijo,$^{1}$ \thanks{e-mail: n.osorio@irya.unam.mx}
O. González-Martín,$^{1}$ 
S.F. Sánchez,$^{2}$
D. Esparza-Arredondo,$^{1}$
J. Masegosa,$^{3}$
\newauthor
C. Victoria-Ceballos,$^{1}$
L. Hernández-García$^{4,5}$
Y. Díaz,$^{5}$
\\
$^{1}$Instituto de Radioastronomía and Astrofisica (IRyA-UNAM), 3-72 (Xangari), 8701, Morelia, Mexico \\
$^{2}$Instituto de Astronomía, Universidad Nacional Autónoma de México, A. P. 70-264, C.P. 04510, México, D.F., Mexico\\
$^{3}$IAA – Instituto de Astrofísica de Andalucía (CSIC), Glorieta de la Astonomía, 18008 Granada, Spain\\
$^{4}$Millennium Institute of Astrophysics (MAS), Nuncio Monse\~nor S\'otero Sanz 100, Providencia, Santiago, Chile \\
$^{5}$Instituto de Física y Astronomía, Facultad de Ciencias, Universidad de Valparaíso, Gran Bretaña 1111, Playa Ancha, Valparaíso, Chile
}

\date{Accepted 2021 December 19. Received 2021 December 16; in original form 2021 March 25.}

\pubyear{2022}

\begin{document}
\label{firstpage}
\pagerange{\pageref{firstpage}--\pageref{lastpage}}
\maketitle

\begin{abstract}

According to theory, the torus of active galactic nuclei (AGN) is sustained from a wind coming off the accretion disk, and for low efficient AGN, it has been proposed that such structure disappears. However, the exact conditions for its disappearance remain unclear. This can be studied throughout the reflection component at X-rays, which is associated with distant and neutral material at the inner walls of the torus in obscured AGN. We select a sample of 81 AGNs observed with NuSTAR with a distance limit of D< 200\,Mpc and Eddington rate $\rm{\lambda_{Edd} \equiv L_{bol}/L_{Edd}<10^{-3}}$. We fit the 3-70\,keV spectra using a model accounting for a partial-covering absorber plus a reflection component from neutral material. We find that the existence of the reflection component spans in a wide range of black-hole mass and bolometric luminosities, with only $\sim$13\%  of our sample (11 sources) lacking of any reflection signatures. These sources fall in the region in which the torus may be lacking in the L-MBH diagram.
For the sources with a detected reflection component, we find that the vast majority of them are highly obscured ($\rm{\log \ N_H > 23}$), with $\rm{\sim 20\%}$ being Compton-thick. We also find an increase on the number of unobscured sources and a tentative increase on the ratio between $\rm{FeK\alpha}$ emission line and Compton-hump luminosities toward $\rm{\lambda_{Edd}=10^{-5}}$, {suggesting that the contribution of the $\rm{FeK\alpha}$ line changes with Eddington ratio. } 
\end{abstract}




\section{Introduction}

\begin{figure*}

    \centering
    \includegraphics[width = 0.67\columnwidth]{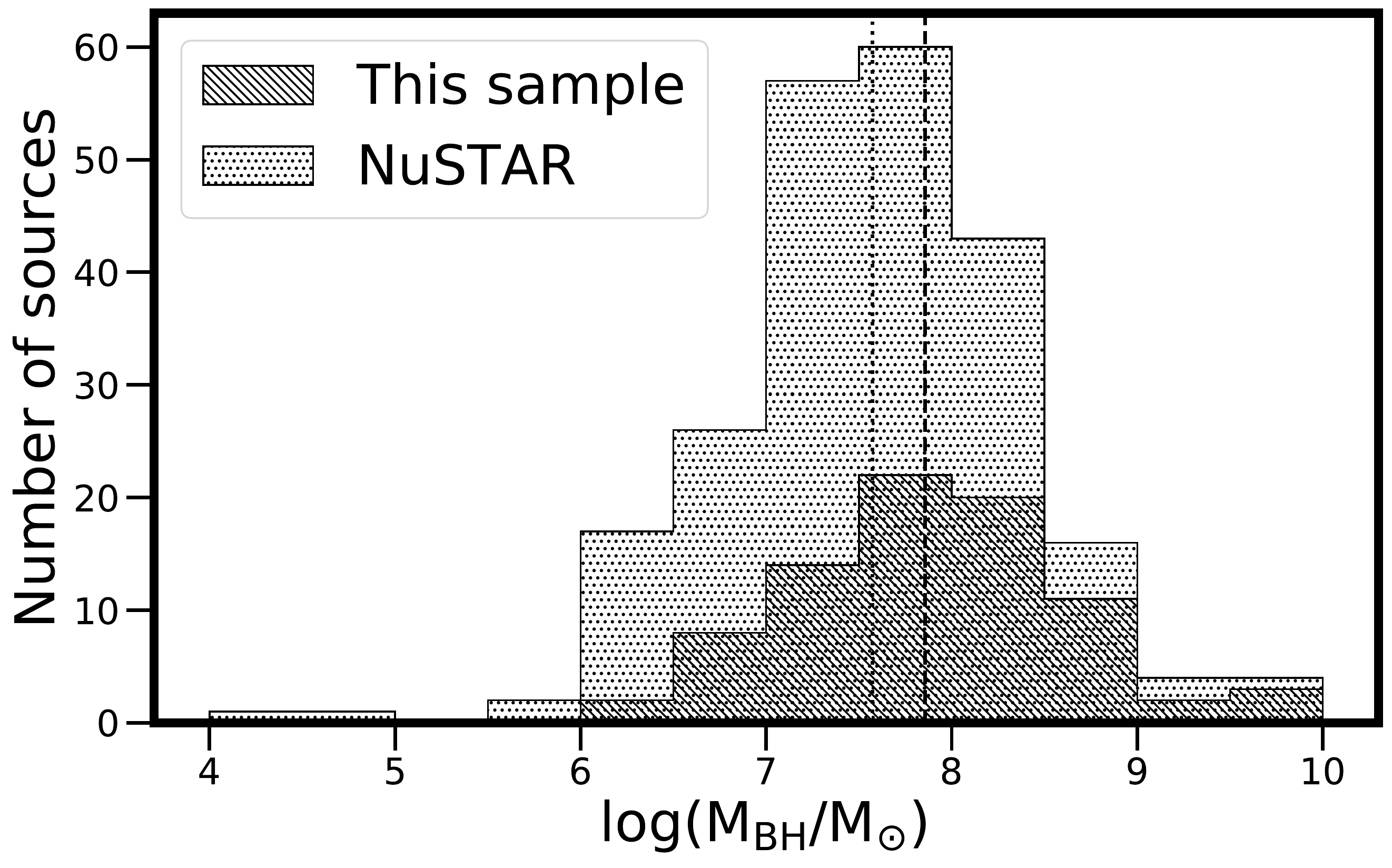}
    \includegraphics[width = 0.635\columnwidth]{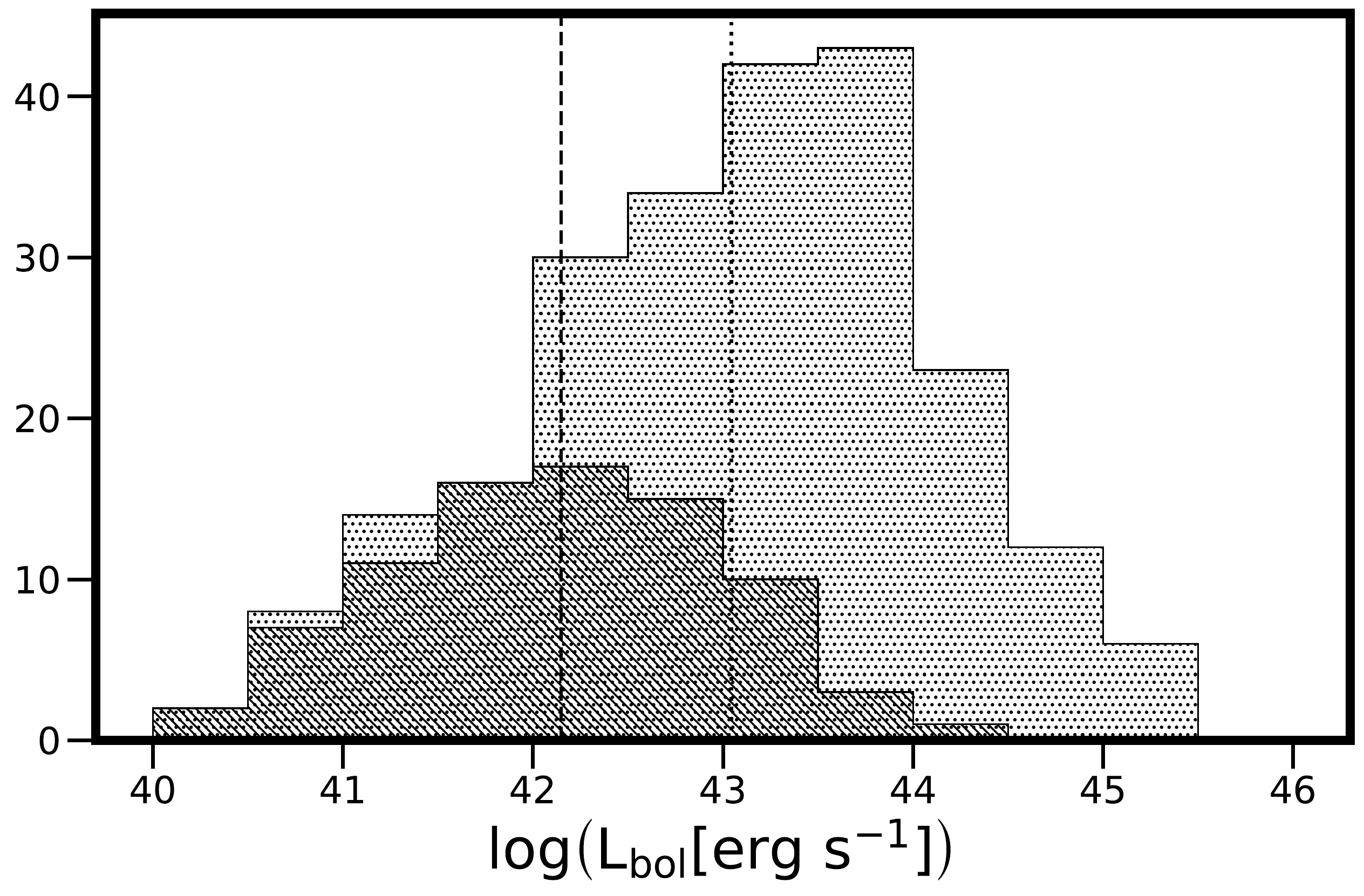}
    \includegraphics[width = 0.635\columnwidth]{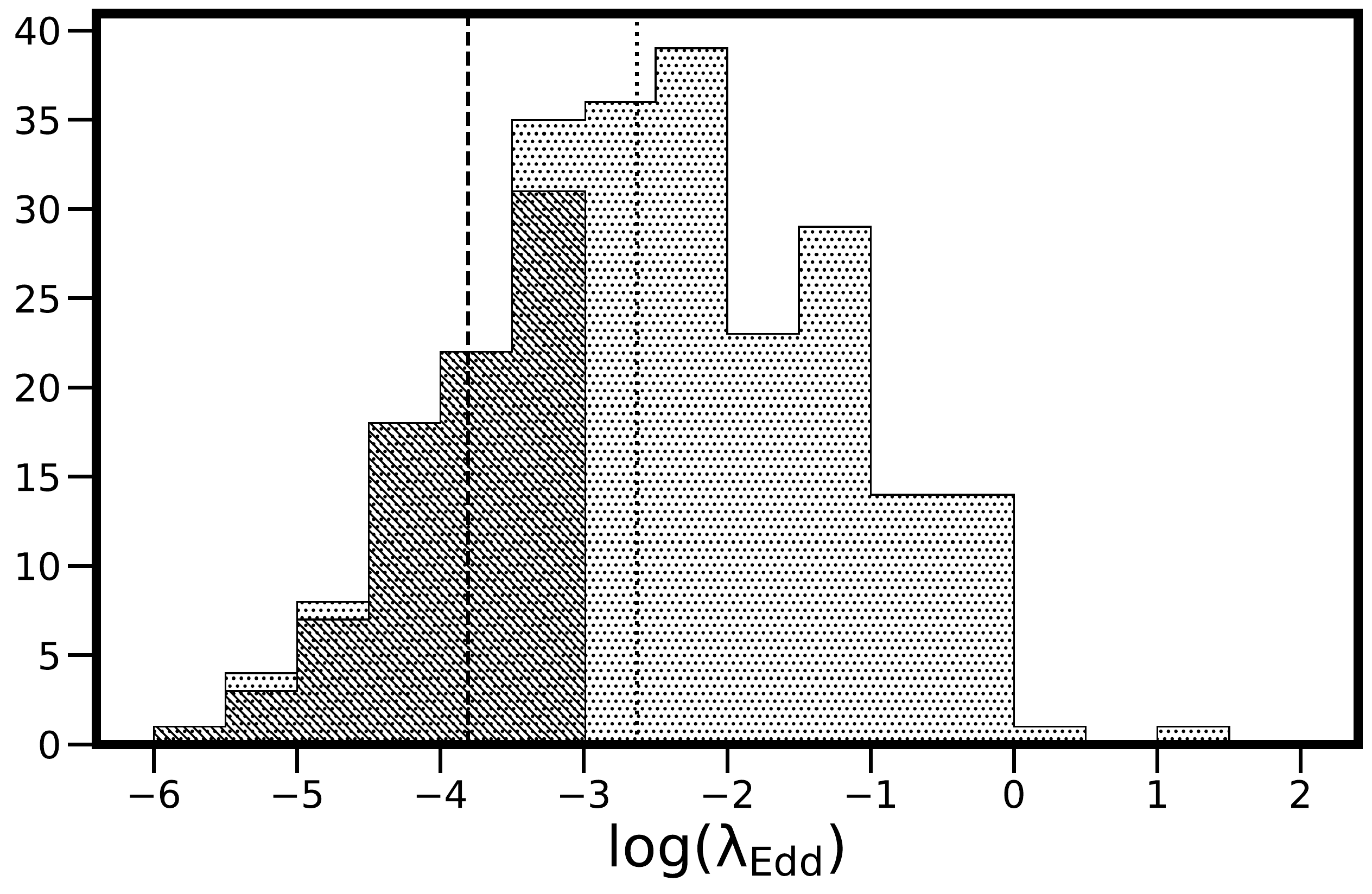}
    
    \caption{Histograms of the distribution of $\rm{M_{BH}}$ (left), $\rm{L_{bol}}$ (center) and $\rm{\lambda_{Edd}}$ (right), for the final sample (dashed, 81 sources) versus the full sample with $\rm{M_{BH}}$ measurements (dotted, 231 sources). The dashed and dotted vertical lines in each panel represent the mean value of the parameter in the X axis for both the final and full samples, respectively. Note that in the left panel, six sources fulfill the accretion criterion but are still excluded from the analysis because they are classified as starburst or because they are at the center of clusters (see text).}
    \label{fig:sample}
\end{figure*}
It has been widely accepted that most galaxies  with bulges host a super massive black-hole (SMBH), sometimes fed by an accretion disk, causing the release of energy in orders of magnitude that can go up to $\rm{L_{bol}\sim 10^{47} \ erg \ s^{-1}}$ \citep[see][for a review]{Netzer15}. Such objects are known as Active Galactic Nuclei (AGN). The Unified Model (UM) of AGN \citep{Antonucci-93, Urry-95} assumes that all AGN have the same components (accretion disk, broad and narrow line regions, a dusty structure-the torus and, in some cases a jet) and that the difference in the spectrum relies on the angle of sight to the observer. This effect may prevent the observer to see the inner parts of AGN and therefore the existence/absence of broad lines in the optical spectrum, leading to a purely observational classification \citep[see][for recent reviews on the topic]{Netzer15,Ramos-Almeida17}. 

The cornerstone of the UM are both the obscuring structure and the inclination angle, as well as the existence of the jet. However, some observations suggest that there are objects lacking the obscuring structure \citep[the so-called true type 2 AGN,][]{Laor-03}. This might be explained under the theoretical prediction that the torus should disappear below a certain luminosity, assuming that both the torus and broad line region (BLR) are formed from a wind coming off the accretion disk \citep{Elvis-00}. In fact, \cite{Elitzur-06} state that both the torus and the BLR should disappear for bolometric luminosities below ($\rm{L_{bol} \sim 10^{42} \ erg \ s^{-1}}$), since the radiation pressure cannot longer counteract the gravity for both structures. Indeed, \citet{Gonzalez-17} using mid-infrared (mid-IR) spectra, have found hints on the torus disappearance in inefficient AGN \citep[see also][]{Gonzalez-15}. Moreover, \cite{Elitzur-09} find that the torus might still be absent for luminous AGN, depending on the efficiency of the accretion disk at feeding the SMBH. Therefore, not only the luminosity of the AGN gives us hints on the existence of the torus, but also the efficiency and density of the wind. Both mid-IR and X-ray studies suggest that the torus is intrinsically different in type-2 AGN than in type-1 AGN \citep[e.g.,][]{Ramos-Almeida09,Ricci-11}, and that the fraction of type-2 sources increases for AGN with Eddington rates $\rm{\lambda_{Edd} \equiv L_{bol}/L_{Edd} \simeq 0.001}$ \citep{Ricci-17}. Indeed, \citet{Khim-17} found a complex behaviour where the number of clouds and covering factor change for high accretion sources (i.e., $\rm{0.01<\lambda_{Edd} < 1}$), forming a ridge-shaped distribution in the luminosity versus black-hole mass ($\rm{M_{BH}}$) diagram. However, the {behaviour} of the torus for low accretion AGN with $\rm{\lambda_{Edd} < 0.001}$, where the torus should evolve to its disappearance, remains an open question.  

Historically, the obscuring matter has been studied through the column density along the line of sight (LOS) that affects the primary X-ray continuum \citep[e.g.][]{Matt03}. Additionally, the torus can also be studied at X-rays because the inner parts of it can reflect the X-ray emission from the corona above the disk, placing a reflection component in the X-ray spectrum. This reflection component has two main signatures: the Compton hump and the FeK$\alpha$ line. Theoretical works have shown that the resulting reflection spectrum depends on the shape and distribution of the reflector, but also on the line-of-sight to the observer, and all of this might have an impact on the luminosity of this component \citep[e.g. see][]{Furui-16}. However, it has not been until recent years that studies aiming to detect and characterize the torus through the reflection component at X-rays have been developed \citep[e.g.][]{Ricci-11, Liu-14, Ricci14, Furui-16, Kawamuro-16}, but also thanks to the unprecedented sensitivity achieved above 10\,keV provided by \textit{NuSTAR} \citep[e.g.,][among others]{Panagiotou-19, Panagiotou-20, Esparza-19, Esparza-21}.

The aim of this study is to trace the {behaviour} of the obscuring material and plausible disappearance of the torus through X-rays for low accretion AGN, compared to what has already been found for more efficient AGN, by using \emph{NuSTAR} data. 
This work is divided as follows: in Section\,\ref{sec:sample} we present the sample and the data reduction. In Section\,\ref{sec:spec-analysis} we present the model and the spectral analysis, while in Section\,\ref{sec:results} we present the results obtained, in Section\,\ref{sec:discussion} we discuss them and finally we present the summary and conclusions of our work in Section\,\ref{sec:conclusions}. Throughout the analysis, we assume a cosmology of $\rm{H_0 = 70 \ km \ s^{-1} \ Mpc^{-1}}$, $\rm{q_0 = 0}$ and $\rm{\Omega_{\lambda_0} = 0.73}$,  and all distances are redshift-independent taken from the NASA Extragalactic Database (NED\footnote{\url{https://ned.ipac.caltech.edu/}}).

\section{Sample and data reduction}
\label{sec:sample}
\subsection{Sample selection}

We search for all available galaxies within a 10 arcmin radius (using SIMBAD database\footnote{\url{http://simbad.u-strasbg.fr/simbad/}}) for all archived pointings within the \textit{NuSTAR} database (until July 2020). We find a total of 1313 galaxies within a distance of 200\,Mpc ($z < 0.05$) to sample the nearby Universe. We then retrieve the activity type, keeping only those galaxies classified as AGN in either NED or SIMBAD databases. We obtain a total of 463 AGN. We download all the data from the HEASARC archive\footnote{\url{https://heasarc.gsfc.nasa.gov/}} and reduce the observations for these 463 AGN (see section \ref{sec:data-red} reduction procedure), keeping only the observations with more than 30\,bins after producing grouped spectra, and at least 10\,bins above 10\,keV. We also remove from the sample those AGN where the \textit{NuSTAR} resolution cannot fully isolate the nuclear spectra for dual AGN. We obtain a total of 301 sources fulfilling these criteria.

We finally search for $\rm{M_{BH}}$. These measurements are needed to obtain an estimate of the Eddington rate. We firstly look for $\rm{M_{BH}}$ calculated via reverberation mapping \cite[e.g.][]{Laor-01, Laor-03,Woo-02, vasudevan-09} and velocity dispersion otherwise \citep[e.g.][]{McKernan-10, vandenbosch-15, vandenbosch-16}, by using the M-$\sigma$ relation, $\rm{M_{BH}}$ \citep[$\rm{\log(M_{BH}/M_{\odot}) = 8.27 + 5.1\log(\sigma/200 \ km \ s^{-1})}$][]{Ferrarese-00, Gebhardt-00}. As for the sources for which the $\rm{M_{BH}}$ is not calculated through the methods reported above, we also search for the $\rm{M_{BH}}$ in different AGN catalogues where the mass are compiled from literature \citep[e.g.][]{Khorunzhev-12,Koss-17, Bar-19}. We obtain measurements for 231 out of the 301 AGN. This sample is now referred to as full sample.

We then fit the spectra of the 231 AGN with an unabsorbed power-law to roughly calculate the bolometric luminosity, $\rm{L_{bol}}$, from X-rays following the $\rm{L_{bol}/L_{(2-10 \ keV)}}$ conversion from \citet{Marconi-04}, where $\rm{L_{bol}}$ is in units of $\rm{L_{\odot}}$:
\begin{multline}
\rm{\log[L_{bol}/L_{(2-10 \ keV)}] = 1.54 \ + 0.24(L_{bol}-12)} \ + \\ 
\rm{0.012(L_{bol}-12)^2 \ - \  0.0015(L_{bol}-12)^3}.
\end{multline} 
\begin{table*}
\centering
\scriptsize
\begin{tabular}{l l l l c l c l l c c l l l l }
\hline 
Name   &   Other name & ra & dec & Redshift & Dist. & $\rm{\log \ M_{BH}}$ & AGN 	&  Galaxy &    Obs. date 				&		Obs ID & Exp. time  & Ext. radius & N. counts & Bins \\ 
&  &  deg & deg & &  Mpc &  & type & type & &  & ks & arcmin \\
       (1)				&	(2)				&		(3)			&		(4)		&	(5)			&	(6)		&  (7)			&	(8)		&	(9)		& (10) & (11) & (12) & (13) & (14) & (15)	 \\ \hline \hline
NGC1052 & PKS0238-084 & 40.26999 & -8.25576 & 0.0048 & 20.6 & 8.4 & L2 & E4 & 2017-01-17 & 60201056002 & 59.75 & 0.5 & 91.27 & 738\\ 
NGC2655 & ARP225 & 133.90721 & 78.22308 & 0.0057 & 24.4 & 8.0 & L2 & Sa-0 & 2016-11-10 & 60160341004 & 15.95 & 2 & 35.30 & 98\\ 
UGC5101 & CGCG289-011 & 143.96539 & 61.35292 & 0.0394 & 168.6 & 8.3 & L1 & S? & 2014-03-21 & 60001068002 & 18.29 & 1 & 24.33 & 51\\ 
M81 & NGC3031 & 148.88822 & 69.06530 & 0.0009 & 3.7 & 7.8 & L1.8 & SAab & 2015-05-18 & 60101049002 & 209.09 & 0.5 & 255.69 & 1252\\ 
NGC3079 & UGC05387 & 150.49085 & 55.67979 & 0.0038 & 16.4 & 7.2 & L2 & SBc & 2013-11-12 & 60061097002 & 21.54 & 0.5 & 29.85 & 128\\ 
UGC5881 & CGCG125-008 & 161.67715 & 25.93155 & 0.0206 & 88.2 & 8.2 & L2 & Sa & 2015-05-17 & 60160409002 & 21.41 & 1 & 37.39 & 169\\ 
NGC3628 & UGC06350 & 170.07091 & 13.58949 & 0.0023 & 9.8 & 7.2 & L2 & SAb & 2017-12-23 & 60371004002 & 50.35 & 1 & 32.46 & 86\\ 
NGC3718 & ARP214 & 173.14522 & 53.06791 & 0.0034 & 14.7 & 8.3 & L1 & SB & 2017-10-27 & 60301031004 & 90.37 & 0.5 & 44.38 & 208\\ 
NGC3998 & UGC06946 & 179.48389 & 55.45359 & 0.0047 & 20.1 & 9.0 & L1.9 & Sa-0 & 2016-10-25 & 60201050002 & 103.94 & 0.5 & 111.02 & 714\\ 
NGC4102 & UGC07096 & 181.59631 & 52.71095 & 0.0046 & 19.5 & 8.2 & L2 & SAB & 2015-11-19 & 60160472002 & 20.57 & 0.5 & 39.14 & 198\\ 
M106 & NGC4258 & 184.74008 & 47.30372 & 0.0017 & 7.3 & 7.5 & L1.9 & SBbc & 2016-01-10 & 60101046004 & 103.62 & 0.5 & 71.20 & 466\\ 
M58 & NGC4579 & 189.43165 & 11.81809 & 0.0043 & 18.4 & 7.9 & L1.9 & Sb & 2016-12-06 & 60201051002 & 117.84 & 0.5 & 116.60 & 738\\ 
NGC5005 & UGC08256 & 197.73463 & 37.05894 & 0.0047 & 20.1 & 8.1 & L1.9 & SABb & 2014-12-16 & 60001162002 & 49.70 & 1 & 27.82 & 40\\ 
NGC6240 & IC4625 & 253.24525 & 2.40099 & 0.0245 & 104.8 & 9.1 & L2 & S0-a & 2014-03-30 & 60002040002 & 30.86 & 0.5 & 55.62 & 444\\ 
MCG+08-31-041 & ARP102B & 259.81038 & 48.98040 & 0.0242 & 103.5 & 8.9 & L1 & E0 & 2015-11-24 & 60160662002 & 22.40 & 1 & 45.28 & 248\\ 
NGC7130 & IC5135 & 327.08121 & -34.95131 & 0.0162 & 69.2 & 7.5 & L2 & Sa & 2016-12-15 & 60261006002 & 42.12 & 0.5 & 23.86 & 63\\ 
NGC7331 & UGC12113 & 339.26709 & 34.41592 & 0.0031 & 13.4 & 7.8 & L2 & Sbc & 2016-05-03 & 40202013002 & 42.97 & 1 & 38.16 & 153\\ 
NGC7479 & UGC12343 & 346.23612 & 12.32288 & 0.0066 & 28.3 & 7.3 & L1.9 & SBbc & 2016-05-12 & 60201037002 & 18.45 & 1 & 27.73 & 82\\

\hline 

\end{tabular}  	
\caption{Observational parameters for the LINER sample. (1) Name of the source; (2) other name (3) right ascension; (4) declination; (5) Redshift; (6) Redshift-independent distance in Mpc; (7) $\rm{\log\left(M_{BH}/M_{\odot}\right)}$; (8) AGN classification; (9) Galaxy type. All galaxy types were retrieved from NED; (10) date of the observation; (11) Observation ID in \textit{NuSTAR}; (12) Exposure time of the observation in ksec; (13) is the extraction radius used in the data reduction; (14) is the number of counts {in the 3-60 keV {\it NuSTAR} band when accounting for the background substraction}; and (15) is the number of bins. In columns (1) and (2), 2MXJ (2MXIJ) is abbreviation for 2MASXJ (2MASXIJ), 2M is the abbreviation for 2MASS and W is the abbreviation for WISE.}
\label{tab:observationalparam1}
\end{table*}

We define low accretion sources as those with $\rm{\lambda_{Edd} < 10^{-3}}$, where $\rm{L_{Edd}}$ is the Eddington luminosity, defined as the maximum luminosity an object can release without losing the thermal equilibrium. { Such defintion is based on the assumption that the transition from the standard disk \citep{Shakura-73}  to an Advection Dominated Accretion disk \citep[ADAF,][]{Narayan-95} occurs around this value \citep[see also,][]{Constantin-09, Gu-09, Younes-11, Gultekin-12, Jang-14, Connolly-16, Hernandez-16}}. This sample contains a total of 87 sources. This corresponds to $\rm{\sim 20\%}$ of the total nearby available AGN observed with \textit{NuSTAR}. 
Four of these sources (NGC\,3310, Mrk\,18, Arp\,299C, and M\,83) are excluded from the sample because although they are classified as AGN in SIMBAD/NED, they are also classified as starbursts. {Finally, we exclude NGC\,4486 due to the strong thermal and jet contribution \citep{DeJong-15}, while we exclude  Centaurus\,A  due to strong jet contribution \citep{Hardcastle-03}. Note that we also check for important thermal/jet contribution for the rest of the sample. Although in some cases there are reported thermal contribution at X-rays \citep[see ][]{Gonzalez-09}, {it does not contribute above 3 keV for the rest of the sources. One consequence , although not the only one, } would be a very steep photon index \citep[(i.e., $\rm{\Gamma} > 2$][]{Nemmen-14} as the one in M\,87 \citep[$\Gamma = 2.8\pm0.4$][]{DeJong-15}, which is not the case in our sample, see section \ref{sec:results}). Our final sample comprises a total of 81 sources. }

Fig.\,\ref{fig:sample} shows the distribution (in logarithmic scale) of $\rm{M_{BH}}$ (left), $\rm{L_{bol}}$ (center) and $\rm{\lambda_{Edd}}$ (right) for our final sample of 81 sources, compared to the full sample of 231 sources with $\rm{M_{BH}}$ estimates.
The mean of the distributions of the final sample for $\rm{M_{BH}}$, $\rm{L_{bol}}$ and $\rm{\lambda_{Edd}}$ are $\rm{<log(M_{BH})>=}$7.9, $\rm{<log(L_{bol})>=}$42.1 and $\rm{log(\lambda_{Edd})=}$-3.8, compared to $\rm{<log(M_{BH})>=}$7.6, $\rm{<log(L_{bol})>=}$43.0 and $<\rm{log(\lambda_{Edd})>=}$-2.6, for the full sample. Thus, our final sample contains lower luminosity AGN with roughly the same range 
of $\rm{M_{BH}}$ values compared to the full sample. Note that this imposes a bias on our sample for Eddington rates above $\rm{log(\lambda_{Edd})>}$-3. Moreover, once we correct by absorption in the line-of-sight, Eddington rates will be higher (see section \ref{sec:luminositycorrection}). Thus, we might lack objects with low obscuration and intrinsically high Eddington rate above $\rm{log(\lambda_{Edd})>}$-2.5. Another source of bias is the fact that our sample relies purely on archival data. For instance, we do not find available \textit{NuSTAR} data for the candidates of torus disappearance in \citet{Gonzalez-17}. Thus, we might also {be missing} the least luminous sources as they might not haven been observed yet. We warn the reader on the fact that our sample is by no means complete, although it does cover a wide range in $\rm{M_{BH}}$ and Eddington rate.

Several works have studied samples of inefficient AGN in different wavelengths {aiming to understand the nature and behaviour of the reflection component} and will be used to compare with our new results along the text. Here we report a brief summary on the matching of our sample with theirs. The largest sample using X-rays is reported by \citet{Gonzalez-09} where they study a sample of 82 inefficient AGN using \textit{Chandra} and \textit{XMM}-Newton data. 
We have 12 sources in common with available \textit{NuSTAR} observations. \citet{Kawamuro-16} also studied a sample of 10 inefficient AGN using \textit{Suzaku} and we have five sources in common with them. On the other hand, \cite{Gonzalez-17} analyzed a sub-sample of those presented by \citet{Gonzalez-09} together with a complementary Seyfert sample (22 sources). We have a total of 39 sources in common with the works mentioned above, which is around 50\% of our final sample. 

Tables\, \ref{tab:observationalparam1} and \ref{tab:observationalparam2} report the observational parameters for the optically classified as low ionization nuclear emission-line regions (LINER, 18 sources) and Seyfert (64 sources) samples, respectively.

\begin{table*}
\centering
\scriptsize

\begin{tabular}{l l l l c l c l l c c l l l l}
\hline 
Name   &   Other name & ra & dec & Redshift & Dist. & $\rm{\log \ M_{BH}}$ & AGN 	&  Galaxy &    Obs. date 				&		Obs ID & Exp. time & Ext. radius & N. counts & Bins\\ 
&  &  deg & deg & &  Mpc &  & type & type & &  & ks & arcmin \\
       (1)				&	(2)				&		(3)			&		(4)		&	(5)			&	(6)		&  (7)			&	(8)		&	(9)		& (10) & (11) & (12) & (13)	& (14) & (15) \\ \hline \hline

NGC253 & ESO474-G029 & 11.88806 & -25.28880 & 0.0008 & 3.2 & 6.9 & S2 & SAB & 2012-09-15 & 50002031004 & 157.65 & 0.5 & 83.01 & 431\\ 
NGC424 & ESO296-G004 & 17.86516 & -38.08345 & 0.0118 & 50.7 & 7.5 & S1 & SB0-a & 2013-01-26 & 60061007002 & 15.48 & 1 & 32.14 & 145\\ 
IC1657 & ESO352-G024 & 18.52924 & -32.65090 & 0.0107 & 45.9 & 7.3 & S2 & SBbc & 2017-01-15 & 60261007002 & 45.16 & 0.5 & 40.39 & 217\\ 
2MXJ01142491-5523497 & NGC0454NED0 & 18.60388 & -55.39705 & 0.0121 & 51.9 & 8.5 & S2 & II & 2016-02-14 & 60061009002 & 24.23 & 1 & 34.16 & 144\\ 
MCG+08-03-018 & 2MXJ01223442+5003180 & 20.64341 & 50.05496 & 0.0204 & 87.4 & 8.4 & S2 & S? & 2014-01-27 & 60061010002 & 31.66 & 1 & 43.88 & 230\\ 
NGC612 & ESO353-G015 & 23.49063 & -36.49328 & 0.0298 & 127.5 & 8.5 & S2 & SA0 & 2012-09-14 & 60061014002 & 16.69 & 0.5 & 31.62 & 158\\ 
Mrk573 & UGC01214 & 25.99074 & 2.34987 & 0.0172 & 73.6 & 7.4 & S2 & S0 & 2018-01-06 & 60360004002 & 32.00 & 1 & 29.73 & 66\\ 
NGC788 & MCG-01-06-025 & 30.27693 & -6.81587 & 0.0136 & 58.3 & 7.7 & S2 & S0-a & 2013-01-28 & 60061018002 & 15.41 & 0.5 & 35.64 & 198\\ 
M77 & NGC1068 & 40.66988 & -0.01329 & 0.0025 & 10.6 & 7.2 & S2 & SAb & 2012-12-18 & 60002030002 & 57.85 & 0.5 & 82.88 & 657\\ 
NGC1106 & UGC02322 & 42.66873 & 41.67158 & 0.0145 & 62.0 & 7.5 & S2 & SA0 & 2019-02-22 & 60469002002 & 18.74 & 1 & 26.56 & 67\\ 
NGC1125 & MCG-03-08-035 & 42.91792 & -16.65111 & 0.0109 & 46.8 & 7.2 & S2 & SAB0 & 2019-06-10 & 60510001002 & 31.74 & 1 & 36.46 & 123\\ 
NGC1142 & UGC02389 & 43.80095 & -0.18355 & 0.0288 & 123.5 & 8.9 & S2 & Spec & 2017-10-14 & 60368001002 & 20.71 & 0.5 & 28.40 & 119\\ 
Mrk1066 & UGC02456 & 44.99415 & 36.82050 & 0.0121 & 51.7 & 7.0 & S2 & SB0 & 2014-12-06 & 60001154002 & 30.08 & 1 & 30.03 & 64\\ 
NGC1194 & UGC02514 & 45.95463 & -1.10375 & 0.0136 & 58.2 & 7.8 & S1.9 & SA0 & 2015-02-28 & 60061035002 & 31.54 & 0.5 & 39.68 & 224\\ 
NGC1229 & ESO480-G033 & 47.04513 & -22.96025 & 0.0363 & 155.4 & 8.3 & S2 & SBb & 2013-07-05 & 60061325002 & 24.92 & 1 & 35.29 & 155\\ 
NGC1320 & MRK0607 & 51.20288 & -3.04226 & 0.0088 & 37.7 & 6.9 & S2 & Sa & 2013-02-10 & 60061036004 & 28.00 & 1 & 34.45 & 120\\ 
NGC1358 & MCG-01-10-003 & 53.41535 & -5.08951 & 0.0134 & 57.5 & 8.1 & S2 & SAB0 & 2017-08-01 & 60301026002 & 50.00 & 0.5 & 39.72 & 206\\ 
NGC1386 & ESO358-G035 & 54.19266 & -35.99927 & 0.0038 & 16.1 & 7.0 & S1 & S0-a & 2016-05-11 & 60201024002 & 26.43 & 1 & 22.29 & 45\\ 
UGC3157 & CGCG468-001 & 71.62399 & 18.46091 & 0.0154 & 66.0 & 8.0 & S2 & SBbc & 2014-03-18 & 60061051002 & 20.09 & 1 & 35.16 & 163\\ 
2MJ05081968+1721481 & CGCG468-002NED01 & 77.08211 & 17.36336 & 0.0175 & 75.0 & 8.6 & S2 & - & 2012-07-23 & 60006011002 & 15.52 & 0.5 & 51.25 & 368\\ 
ESO5-4 & IRAS06220-8636 & 91.42384 & -86.63195 & 0.0060 & 25.9 & 7.6 & S2 & Sb & 2015-11-10 & 60061063002 & 24.70 & 1 & 33.74 & 109\\ 
NGC2273 & UGC03546 & 102.53614 & 60.84580 & 0.0068 & 29.0 & 7.0 & S2 & SBa & 2014-03-23 & 60001064002 & 23.23 & 1 & 36.69 & 150\\ 
UGC3601 & CGCG204-032 & 103.95638 & 40.00031 & 0.0171 & 73.3 & 8.6 & S1.5 & S? & 2019-01-06 & 60160278002 & 19.67 & 1 & 45.70 & 257\\ 
ESO428-14 & MCG-05-18-002 & 109.13003 & -29.32469 & 0.0054 & 23.2 & 7.0 & S2 & SA0 & 2015-01-11 & 60001152002 & 40.25 & 0.5 & 29.46 & 111\\ 
2MXJ07561963-4137420 & WAJ075619.61-413742.1 & 119.08182 & -41.62835 & 0.0210 & 90.1 & 8.0 & S2 & - & 2014-07-29 & 60061076002 & 22.75 & 2 & 38.87 & 62\\ 
NGC2788A & ESO060-G024 & 135.66418 & -68.22683 & 0.0144 & 61.6 & 8.7 & S2 & Sb & 2019-06-14 & 60469001002 & 27.58 & 0.5 & 27.18 & 99\\ 
IC2560 & ESO375-G004 & 154.07795 & -33.56381 & 0.0078 & 33.4 & 7.2 & S2 & SBbc & 2014-07-16 & 50001039004 & 49.56 & 1 & 35.56 & 99\\ 
NGC3147 & UGC05532 & 154.22347 & 73.40065 & 0.0092 & 39.6 & 8.7 & S2 & SAbc & 2015-12-27 & 60101032002 & 49.26 & 0.5 & 48.47 & 293\\ 
NGC3393 & ESO501-G100 & 162.09778 & -25.16203 & 0.0125 & 53.6 & 7.5 & S2 & SBab & 2013-01-28 & 60061205002 & 15.66 & 0.5 & 25.00 & 87\\ 
2MXJ11055897+5856456 & CGCG291-028 & 166.49593 & 58.94603 & 0.0271 & 116.0 & 8.4 & S2 & - & 2019-03-26 & 60160420002 & 15.77 & 2 & 38.17 & 134\\ 
NGC3621 & ESO377-G037 & 169.56792 & -32.81260 & 0.0016 & 6.7 & 6.8 & S2 & SA & 2017-12-15 & 60371002002 & 30.78 & 2 & 41.09 & 100\\ 
NGC3786 & UGC06621 & 174.92714 & 31.90943 & 0.0118 & 50.6 & 7.5 & S1.8 & SAB & 2014-06-09 & 60061349002 & 21.99 & 2 & 36.08 & 103\\ 
IC751 & UGC06972 & 179.71915 & 42.57034 & 0.0312 & 133.6 & 8.6 & S2 & Sb & 2013-02-04 & 60061217004 & 52.02 & 0.5 & 25.72 & 63\\ 
M88 & NGC4501 & 187.99673 & 14.42041 & 0.0042 & 18.0 & 7.5 & S2 & SAb & 2018-01-26 & 60375002002 & 62.77 & 1 & 27.45 & 55\\ 
IC3639 & ESO381-G008 & 190.22015 & -36.75585 & 0.0109 & 46.8 & 6.9 & S2 & SBbc & 2015-01-09 & 60001164002 & 58.73 & 0.5 & 25.34 & 51\\ 
NGC4785 & ESO219-G004 & 193.36382 & -48.74915 & 0.0116 & 49.6 & 8.1 & S2 & SAB & 2014-08-20 & 60001143002 & 48.83 & 0.5 & 40.05 & 208\\ 
Mrk231 & UGC08058 & 194.05931 & 56.87368 & 0.0422 & 180.6 & 8.4 & S2 & Sc & 2017-10-19 & 80302608002 & 82.06 & 0.5 & 46.30 & 242\\ 
NGC4941 & PGC045165 & 196.05461 & -5.55160 & 0.0033 & 14.2 & 6.9 & S2 & SABa & 2016-01-19 & 60061236002 & 20.66 & 1 & 32.48 & 119\\ 
NGC4939 & MCG-02-33-104 & 196.05970 & -10.33953 & 0.0085 & 36.4 & 7.9 & S2 & Sbc & 2017-02-17 & 60002036002 & 22.04 & 0.5 & 36.10 & 181\\ 
NGC4945 & ESO219-G024 & 196.36366 & -49.46790 & 0.0010 & 4.2 & 6.3 & S2 & SBc & 2013-06-15 & 60002051004 & 54.62 & 0.5 & 129.04 & 1565\\ 
MCG-03-34-064 & PGC046710 & 200.60202 & -16.72836 & 0.0200 & 85.6 & 8.1 & S1.8 & SB? & 2016-01-17 & 60101020002 & 78.50 & 0.5 & 80.69 & 727\\ 
NGC5135 & ESO444-G032 & 201.43358 & -29.83368 & 0.0015 & 6.3 & 7.6 & S2 & Sab & 2015-01-14 & 60001153002 & 33.36 & 1 & 34.28 & 88\\ 
M51 & ARP085 & 202.46957 & 47.19526 & 0.0017 & 7.3 & 6.6 & S2 & - & 2017-03-17 & 60201062003 & 163.06 & 0.5 & 41.19 & 99\\ 
NGC5252 & UGC08622 & 204.56613 & 4.54265 & 0.0195 & 83.6 & 8.9 & S1.9 & S0 & 2013-05-11 & 60061245002 & 19.01 & 0.5 & 65.88 & 523\\ 
NGC5283 & UGC08672 & 205.27395 & 67.67222 & 0.0106 & 45.3 & 7.7 & S2 & S0 & 2018-11-17 & 60465006002 & 33.02 & 0.5 & 42.87 & 234\\ 
NGC5347 & UGC08805 & 208.32416 & 33.49083 & 0.0050 & 21.6 & 6.8 & S2 & Sab & 2015-01-16 & 60001163002 & 47.30 & 1 & 32.04 & 54\\ 
NGC5643 & ESO272-G016 & 218.16991 & -44.17461 & 0.0027 & 11.4 & 7.0 & S2 & Sc & 2014-05-24 & 60061362002 & 22.46 & 1 & 36.54 & 146\\ 
NGC5695 & UGC09421 & 219.34223 & 36.56783 & 0.0125 & 53.5 & 7.7 & S2 & SBb & 2018-01-16 & 60368004002 & 41.61 & 1 & 27.60 & 43\\ 
NGC5728 & MCG-03-37-005 & 220.59970 & -17.25317 & 0.0071 & 30.3 & 7.8 & S2 & SABa & 2013-01-02 & 60061256002 & 24.36 & 0.5 & 52.02 & 402\\ 
NGC5899 & UGC09789 & 228.76355 & 42.04985 & 0.0090 & 38.6 & 8.7 & S2 & SABc & 2014-04-08 & 60061348002 & 23.88 & 0.5 & 54.44 & 383\\ 
MCG+14-08-004 & CGCG367-009 & 244.83058 & 81.04650 & 0.0239 & 102.4 & 9.8 & S2 & - & 2014-12-21 & 60061270002 & 29.76 & 1 & 38.54 & 165\\ 
ESO137-34 & 2MXJ16351411-5804481 & 248.80881 & -58.08003 & 0.0077 & 33.0 & 8.0 & S2 & SAB0 & 2016-06-07 & 60061272002 & 18.55 & 0.5 & 28.69 & 111\\ 
2MXJ16504275+0436180 & NGC6230NED01 & 252.67813 & 4.60508 & 0.0321 & 137.3 & 9.8 & S2 & - & 2017-02-06 & 60061273002 & 21.03 & 0.5 & 56.94 & 409\\ 
2MXIJ1802473-145454 & WAJ180247.38-145454.8 & 270.69708 & -14.91528 & 0.0034 & 14.6 & 7.8 & S1 & - & 2016-05-01 & 60160680002 & 19.96 & 0.5 & 80.72 & 595\\ 
2MXJ18305065+0928414 & LEDA1365707 & 277.71098 & 9.47830 & 0.0194 & 83.2 & 8.4 & S2 & - & 2015-11-15 & 60061285002 & 22.72 & 1 & 39.73 & 183\\ 
IC4995 & ESO186-G034 & 304.99574 & -52.62192 & 0.0161 & 68.9 & 7.2 & S2 & SA0 & 2019-06-03 & 60360003002 & 34.00 & 1 & 32.33 & 70\\ 
NGC6921 & UGC11570 & 307.12018 & 25.72339 & 0.0145 & 62.0 & 8.4 & S2 & SA0 & 2013-05-18 & 60061300002 & 19.52 & 1 & 33.67 & 117\\ 
NGC7213 & ESO288-G043 & 332.31754 & -47.16669 & 0.0051 & 22.0 & 7.7 & S1.5 & Sa & 2014-10-05 & 60001031002 & 101.62 & 0.5 & 160.49 & 965\\ 
NGC7319 & UGC12102 & 339.01501 & 33.97588 & 0.0109 & 46.7 & 7.3 & S2 & SBbc & 2017-09-27 & 60261005002 & 41.88 & 0.5 & 30.20 & 88\\ 
UGC12282 & CGCG532-004 & 344.73035 & 40.93221 & 0.0169 & 72.4 & 9.8 & S1.9 & Sa & 2019-11-18 & 60160812002 & 28.56 & 1 & 27.50 & 43\\ 
NGC7582 & ESO291-G016 & 349.59842 & -42.37057 & 0.0049 & 21.2 & 7.6 & S1.5 & SBab & 2016-04-28 & 60201003002 & 48.49 & 0.5 & 123.14 & 1100\\ 
2MXJ23252420-3826492 & IRAS23226-3843 & 351.35078 & -38.44700 & 0.0359 & 153.8 & 8.2 & S1 & - & 2017-06-11 & 80101001002 & 96.61 & 0.5 & 30.57 & 80\\ 
NGC7674 & UGC12608 & 351.98624 & 8.77895 & 0.0174 & 74.5 & 7.6 & S2 & SAbc & 2014-09-30 & 60001151002 & 52.00 & 0.5 & 36.08 & 161\\

\hline 

\end{tabular}  	
\caption{Observational parameter for the Seyfert sample. Columns are as in Tab.\,\ref{tab:observationalparam1}.}
\label{tab:observationalparam2}
\end{table*}

\subsection{Data reduction}
\label{sec:data-red}
All \emph{NuSTAR} \citep{Harrison-13} data were reduced using the data analysis software \emph{NuSTARDAS} v.1.4.4 distributed by the High Energy Astrophysics Archive Research Center (HEASARC). The calibrated, cleaned and screened event files for both FPMA and FMPB focal plane modules were generated using the {\sc nupipeline} task (CALDB 20160502). We automatically extracted three circular apertures with radius of 0.5, 1, and 2\,arcsec centred at the NED position of the source. For each of them we produced eight backgrounds around the target. These backgrounds are located at a distance of three times the aperture radius of the target (i.e. 1.5, 3, or 6 arcsec) with position angles of 0, 45, 90, 135, 180, 225, 270, and 315 degrees. We produced in this way 24 spectra for each target. The Redistribution Matrix and Auxiliary Response files (RMF and ARF)  were produced with the {\sc nuproducts} package available in \emph{NuSTARDAS}. We chose the spectrum that maximises the S/N in the 3-70 keV band, avoiding off-nuclear contributors. Tables\,\ref{tab:observationalparam1} and \ref{tab:observationalparam2} provide the best extraction radius for each target. Note the optimal extraction radius is $\sim$1\,arcmin ($D \simeq[1-52]$ kpc) for the vast majority of the objects, due to the relative faintness of our sources. 

We then performed a binning method with a minimum S/N ratio of 3, using the  {\sc ftgrouppha} task within FTOOLS. The grouping using a minimum S/N ratio is particularly relevant for faint sources which are probably background dominated above $\rm{\sim}$25 keV. Note that we also performed the standard grouping method accounting for a minimum of 30 counts per bin (the {\sc grppha} task) finding similar results in the spectral parameters of the sources. However, the grouping performed with a minimum S/N ratio clearly improves the error bars at high energies for faint sources as expected.

\section{Spectral Analysis}
\label{sec:spec-analysis}

\subsection{Baseline models}
{ In this section we present four models used in the analysis to fit the spectra, which are fitted  using {\sc xspec} \citep[][version 12.10.0]{Keith-96}\footnote{ \url{http://cxc.heasarc.gsfc.nasa.gov/docs/xanadu/xspec/}}.}  Throughout the analysis, we use the $\chi^2$ statistics. 
We begin the analysis by modelling the intrinsic continuum with a power-law affected by a patchy absorber, in what is known as the partial covering scenario. We add two Gaussian profiles centred at 6.7 and 6.97\,keV, to account for the Fe\,{\sc xxv} and Fe\,{\sc xxvi} ionized emission lines  as they are the most common ionized lines present at hard X-rays for AGN. In the software terminology, this model has the following form:
\begin{multline}
M_{1} =  {\tt phabs_{\rm{Gal}}}
(({\tt zphabs_{\rm{intr}}}*{\tt zpowerlw}) +  {\tt ct}*{\tt zpowerlw} \\
+ {\tt zgauss_{6.7 \ keV}} + {\tt zgauss_{6.97 \ keV}})
\end{multline}

\noindent where $({\tt zphabs_{\rm{intr}}}*{\tt zpowerlw}) + {\tt ct}*{\tt zpowerlw}$ represents the partial-covering scenario, in which the first power-law component is associated with the intrinsic continuum absorbed by the material along the LOS to the observer and the second power-law component is the scattered emission that reaches the observer. The parameters for both power-laws are linked, including the normalization, to simulate the partial-covering scenario. The free parameters for this model are the column density ($\rm{N_H}$), the constant associated with the covering factor (as $\rm{ct=1-f_{cov}}$), the photon index ($\rm{\Gamma_{pl}}$), and the normalization of the power-law.
The free parameters of the Gaussian components, emulating the emission lines are the center, width and normalization. The widths of the Gaussian components are fixed to be narrow (0.1 keV\footnote{The \textit{NuSTAR} spectral resolution is 400 eV at 6 keV, see \\ \url{https://heasarc.gsfc.nasa.gov/docs/nustar/nustar_tech_desc.html}}) and the centers are fixed to 6.7 and 6.97 keV, respectively. 

For the second and third models we include an extra reflection component as follows:
\begin{multline}
M_{2} =  M_{1} + {\tt phabs_{\rm{Gal}}}({\tt pexrav} + {\tt zgauss_{6.4 \ keV}})  
\end{multline}
\begin{multline}
M_{3} =  M_{1} + {\tt phabs_{\rm{Gal}}}({\tt pexmon})
\end{multline}

For $\rm{M_2}$, we choose as reflection component the {\tt pexrav} model \citep{Magdziarz-95}.  This model assumes Compton reflection from neutral X-ray photons in an optically thick material with plane-parallel geometry. This model has as free parameters the photon index,  high energy cutoff, relative reflection {(R)}, metal and iron abundances, inclination angle, and normalization. We have set the photon index of {\tt pexrav} ($\rm{\Gamma_{pex}}$) to be the same as that of the power-law, assuming that the reflection occurs in the AGN such that both emissions are correlated. The energy cutoff is fixed to 300\,keV assuming this cutoff to happen above the {\it NuSTAR} energy range. Note that although several works \citep[e.g.][]{Molina-19, Younes-19, Ezhikode-20} have aimed to find this cutoff in AGN spectra, it has only been possible for very few objects, and in most cases, it happens well above the \textit{NuSTAR} range \citep[e.g][]{Balokovic-20}. Therefore, we do not expect for this parameter to be constrained nor to affect the overall analysis. {We have set R to $-1$} for {\tt pexrav} to account for the reflection component only, without contribution from the intrinsic power-law continuum. Thus, this reflection model with the relative reflection set to $-1$ plus the power-law component is equivalent to the {\tt pexrav} allowing the relative reflection to vary. However, we used with the relative reflection to $-1$ in order to separate both the intrinsic continuum and the reflection components.

The rest of the parameters are set to their default values except for the normalization. Note that in this model version, the addition of an extra Gaussian profile is necessary to emulate the FeK$\rm{\alpha}$ line, since {\tt pexrav} assumes reflection from a neutral material, and does not account for this line. In this model, the width of the line at 6.4\,keV is also fixed to 0.1\,keV.

For $\rm{M_3}$, we include the reflection component using {\tt pexmon} \citep{Nandra-07}. This model is an updated version of {\tt pexrav} that includes fluorescence lines such as FeK$\rm{\alpha}$, FeK$\rm{\beta}$ and NiK$\rm{\alpha}$. This model also accounts for the Compton shoulder, by assuming a Gaussian line at 6.315\,keV with a width of 0.35\,keV. Additionally, it assumes a connection between the equivalent width (EW) of the FeK$\rm{\alpha}$ line and that of the Compton shoulder. This reflection model has as free parameters the photon index ($\rm{\Gamma_{pex}}$), relative reflection, cutoff energy, metal and iron abundances, inclination angle, and normalization. Similar to $\rm{M_2}$, the {parameter R } is set to $-1$, the energy cutoff is set to 300\,keV and the other parameters, except the normalization, are set to their default values.

The difference between $\rm{M_2}$ and $\rm{M_3}$ relies on the fact that {\tt pexmon} accounts for the abundance of iron in the medium through the inclusion of the FeK$\rm{\alpha}$, FeK$\rm{\beta}$ and NiK$\rm{\alpha}$ lines. Indeed, these lines are correlated with the Compton-reflection as they are fixed to a fraction of the Compton-shoulder, and they also depend on the photon index of the intrinsic continuum. {\tt pexmon} also accounts for the Compton-shoulder, while {\tt pexrav} is a simpler model not accounting for such features. Thus, {\tt pexmon} already incorporates the $\rm{FeK\alpha}$ line at 6.4\,keV while we add this line using a Gaussian profile in $\rm{M_2}$. We use $\rm{M_2}$ to calculate the EW and luminosity of the 6.4\,keV line, with the normalization and error of the line component and the tools {\sc equivalent} and {\sc luminosity} within {\sc xspec}. $\rm{M_2}$ also allows us to isolate the effect of the line from that of the Compton shoulder. Note that both models are equivalent in the analysis. However, since {\tt pexmon} self-consistently accounts for {both the continuum and line emission due to the reprocessing in neutral, distant material}, the rest of the analysis is based on this model. We include $\rm{M_1}$ to study the detection of the reflection component in our sample through the f-statistic test when comparing models (see below). Note also that regardless of the model, we account for Galactic absorption and redshift by using the {\sc nh} tool within {\sc ftools} \citep[retrieved from NED\footnote{\url{https://ned.ipac.caltech.edu}} and fixed to the HI maps of][]{Kalberla-05}.

We also use a fourth model that accounts for reflection dominated sources, i.e., sources for which the intrinsic continuum is completely covered by the reflection component in the available X-ray data. 
This model has the following form:
\vspace{-0.2cm
\vspace{-0.2cm}}
\begin{multline}
M_{4} =  {\tt phabs_{\rm{Gal}}}*{\tt pexmon}
\end{multline}

\begin{figure*}
    \centering
    \includegraphics[width = 1\linewidth, height = 0.8\linewidth]{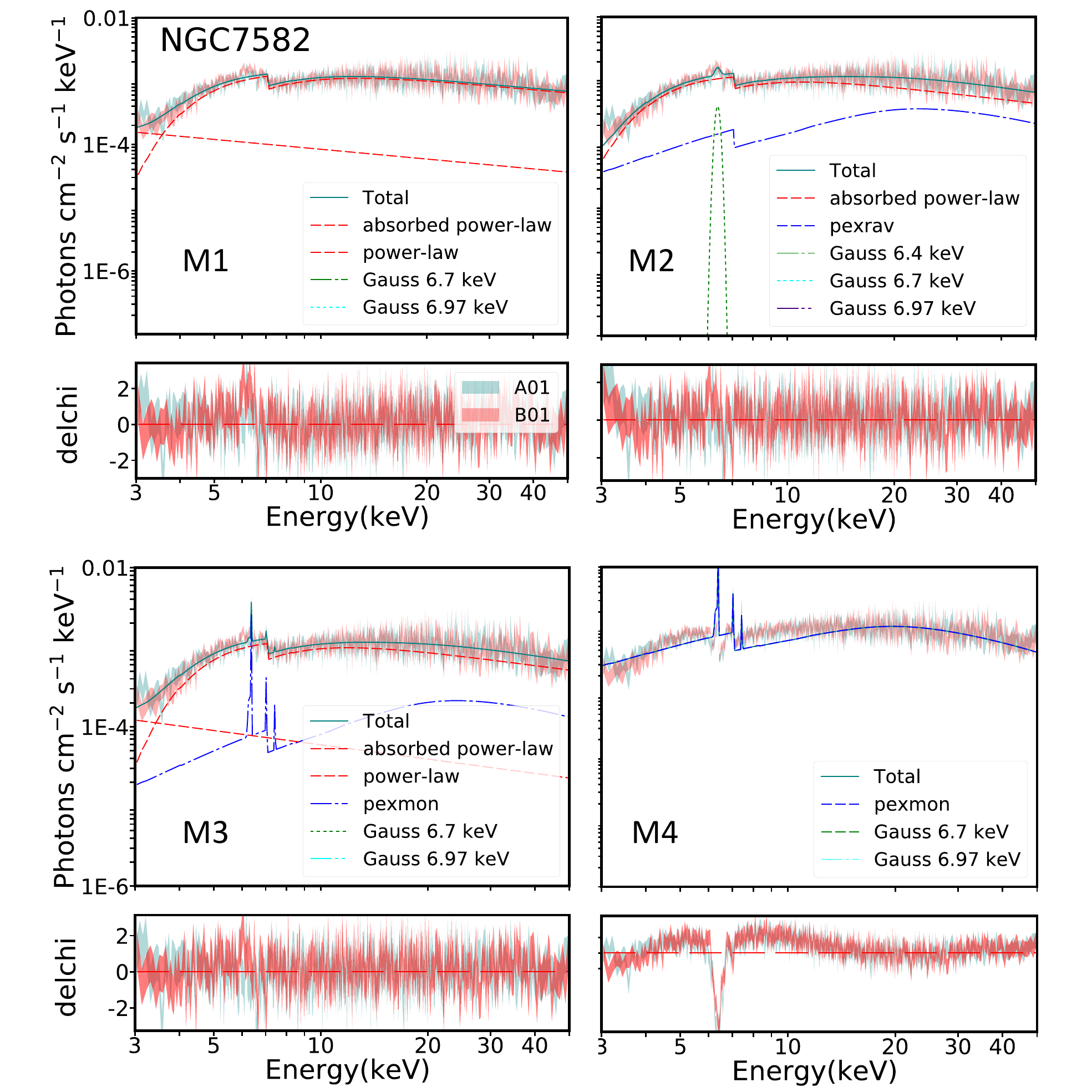}

    \caption{Spectral fits for $\rm{M_1}$ (upper left), $\rm{M_2}$ (upper right), $\rm{M_3}$ (lower left) and $\rm{M_4}$ (lower right) for NGC7582. Each color represents each of the components as follows: teal is the total spectrum, red is the partial-covering component, blue are pexmon (left) and pexrav (right) components, while green, cyan and purple are the extra lines in each model version. (see section \ref{sec:spectral-fitting})}
    \label{fig:models}
\end{figure*}

This analysis does not use sophisticated models as {\sc borus} \citep{Balokovic-18} or {\sc mytorus} \citep{Murphy-09} for neutral torus reflection because the faintness of the sources in our sample may not allow us to restrict geometrical parameters for the majority of them. {Indeed, we also tested the {\sc mytorus} model, but the S/N of the data were too low for parameters to be restriced in most cases.} Instead, we prefer a homogeneous analysis of the strength and general characteristics of the reflection component in our sample. This is sufficient to study the dependence of the existence of the reflection component with the AGN power, which is one of the main purposes of this analysis. Furthermore, it also avoids the use of ionized disk reflection components, such as {\sc relxill} \citep{Dauser-14} because disk reflection has been found only for a small fraction of AGN, all of them in the high accretion regime \citep{Esparza-21}. Finally, this spectral analysis lacks the inclusion of the Compton scattering, which contributes specially for Compton-thick\footnote{\url{https://ned.ipac.caltech.edu/level5/March04/Comastri/Comastri_contents.html}} (CT) sources. However, the available models within {\sc xspec} (e.g., {\sc cabs}) do not properly account for it as they assume an inadequate cross-section in the modelling \citep[cf.,][and {\sc mytorus} manual\footnote{\url{http://mytorus.com}}.]{Murphy-09}. {Indeed, we also try the scenario in which {\sc cabs} is included but we find the intrinsic luminosity to be unrealistically high in some cases (see Appendix \ref{Appendix-A}). Note that the exclusion of such a component leads to a systematic uncertainty, which is translated into an underestimation of the intrinsic luminosities of CT sources. However, our baseline model has been widely used in several works aimed to find general properties of AGN and characterize the reflection component, even in CT objects \citep[e.g.,][]{Kawamuro-16, Panagiotou-19, Panagiotou-20, Kang-20}}. Thus, the baseline model used in this work will lead to robust results that can be easily compared with previous works \citep[e.g.,][]{Kawamuro-16, Annuar-15, Annuar-17, Annuar-20}. 

\subsection{Spectral fitting}
\label{sec:spectral-fitting}
We start the analysis by fitting the spectra to $\rm{M_1}$. We then fit the spectra to $\rm{M_2}$, and obtain the EW and the luminosity of the FeK$\rm{\alpha}$ line as explained above. We then fit the spectra to $\rm{M_3}$ and test if reflection is required by the data using the f-test tool within {\sc xspec}. If the f-test throws a probability value below $10^{-3}$, then the reflection component is required by the data. For those objects for which the reflection component is required, we also test the statistical need for a power-law (i.e., we fit the spectra to $\rm{M_4}$). Objects for which such component is not statistically needed will be referred as reflection dominated.

We also estimate the reflection fraction, defined in this work as $\rm{C_{ref} = L_{ref}/ L_{cont}}$, where $\rm{L_{ref}}$ is the luminosity of the reflection component and $\rm{L_{cont}}$ is the luminosity of the intrinsic continuum, modelled with a power-law, with both luminosities calculated in the 3-70\,keV energy range. They are computed using the {\sc cluminosity} command within {\sc xspec}. We estimate the 1-$\sigma$ errors for the best-fit parameters. Errors in the luminosities of the different components, as well as in the EW of the FeK$\rm{\alpha}$ line are presented as follows: if the 3-$\sigma$ errors are well constrained, we present the 1-$\sigma$ error; upper/lower limits correspond to the 3-$\sigma$ estimates otherwise. This makes the analysis more conservative on the luminosity limit of the reflection component and the FeK$\rm{\alpha}$ line for those sources where the reflection component or the FeK$\rm{\alpha}$ line are not recovered by the data since it gives 99.6\% the upper limit of the distribution. As for the reflection fraction, we perform MCMC simulations to recover its value and error as follows: we choose a random number within a normal distribution centered at the actual value whenever the luminosity is well constrained. However, when there is an upper/lower limit of the luminosity, we choose a random number drawn from an uniform distribution with limits zero and the upper/lower limit. We then calculate the ratio and repeat the same procedure 1000 times. Finally, we obtain the median for the central value and the $25^{\rm{th}}$ and $75^{\rm{th}}$ quartiles as the error range. Whenever the upper limit is above 1 or the lower limit is below 0, we report the value as a lower or upper limit, respectively.

Fig.\,\ref{fig:models} shows the best fit for $\rm{M_1}$ (upper left), $\rm{M_2}$ (upper right), $\rm{M_3}$ (lower left) and $\rm{M_4}$ (lower right) for NGC\,7582 as an example of the resulting spectral fits. For this particular object, we obtain for $\rm{M_1}$ a $\rm{\chi^2/dof = 699/675}$, $\rm{M_2}$ a $\rm{\chi^2/dof = 655/674}$ for while for $\rm{M_3}$ we obtain $\rm{\chi^2/dof = 650/674}$, for which the f-test gives a value of $\rm{10^{-12}}$. Indeed, in the upper left panel of the figure, it is possible to see the poor fit and large residuals in particular around the $\rm{FeK\alpha}$ line, indicating the need of an extra component to fit the spectrum. Also note that although upper right and lower left panels show different models, they both fit the data well, by accounting for the reflection component, mostly noticeable through the FeK$\rm{\alpha}$ line. On the other hand, $\rm{M_4}$ fits the data poorly with a $\rm{\chi^2/dof = 3115.7/677}$ (see lower right panel). Indeed, although required, the line accounts only for $\sim 0.9\%$ of the reflection component, while the latter only accounts for $\sim 20\%$ of the 3-70\,keV primary continuum luminosity. This is an example where the existence of the $\rm{FeK\alpha}$ emission line, the f-test when comparing with the partial-covering intrinsic continuum model, and the luminosity detection limit, point out to the existence of a reflection component for this source. 

\section{Results}\label{sec:results}
\begin{table*}
\centering
\scriptsize
\begin{tabular}{l l l l l l l l l l l l l}
\hline 
Name   &  $\rm{\chi^2/dof}$ &   $\rm{\log \ NH}$ & $\rm{f_{cov}}$ & $\rm{\Gamma}$ & $\rm{EW}$ & $\rm{C_{ref}}$ & \multicolumn{2}{c}{$\rm{\log \ L_{(2-10) \ keV}  }$}  & \multicolumn{2}{c}{$\rm{\log \ L_{(3-70) \ keV}  }$} & ftest \\
\cline{8-9} \cline{10-11} \\
& & & & & & & $\rm{continuum}$ & $\rm{reflected}$ & $\rm{continuum}$ & $\rm{reflected}$  \\  

   &   &  $\rm{cm^{-2}}$  &  &   & $\rm{eV}$ & & $\rm{erg \ s^{-1}}$  & $\rm{erg \ s^{-1}}$& $\rm{erg \ s^{-1}}$   & $\rm{erg \ s^{-1}}$  \\
       (1)				&	(2)				&		(3)			&		(4)		&	(5)			&	(6)		&  (7)			&	(8)		&	(9) & (10) & (11) & (12)	 \\ \hline \hline

NGC1052 & $744.71/734$ &  ${23.15}_{-0.03}^{+0.03}$ & $1*$ & ${1.78}_{-0.04}^{+0.04}$  & $139.30 \pm 27.4 $ &  ${0.15}_{-0.06}^{+0.06}$  & $41.73 \pm 0.01 $ & $40.21 \pm 0.15$ & $42.14 \pm 0.01$ & $41.32\pm0.15$ &\checkmark \\
NGC2655 & $84.87/94$ &  ${23.41}_{-0.14}^{+0.12}$ & $1*$ & ${2.05}_{-0.15}^{+0.17}$  & $368.43 \pm 105. $ &  ${0.56}_{-0.26}^{+0.35}$  & $41.32 \pm 0.10 $ & $40.30 \pm 0.17$ & $41.59 \pm 0.07$ & $41.23\pm0.17$ &x \\
UGC5101 & $54.98/47$ &  ${23.84}_{-0.06}^{+0.06}$ & $1*$ & $<1.26$  & $< 451.$ & $< 1 $ & $42.61 \pm 0.09 $ & $< 41.7$ & $43.45 \pm 0.08$ & $< 43.4$ &x \\
M81 & $1372.00/1246$ &  ${22.16}_{-0.09}^{+0.13}$ & - & ${1.94}_{-0.01}^{+0.01}$  & $33.78 \pm 9.45 $ & $< 0.024 $ & $40.54 \pm 0.01 $ & $< 38.2$ & $40.88 \pm 0.15$ & $< 39.2$ &x \\
NGC3079\textsuperscript{CT} & $114.37/123$ &  ${24.41}_{-0.05}^{+0.03}$ & ${0.98}_{-0.01}^{+0.01}$ & ${1.68}_{-0.15}^{+0.16}$  & $479.64 \pm 210. $ &  ${0.10}_{-0.05}^{+0.06}$  & $41.59 \pm 0.13 $ & $39.83 \pm 0.20$ & $42.06 \pm 0.05$ & $41.01\pm0.20$ &\checkmark \\
UGC5881 & $136.07/163$ &  ${23.43}_{-0.36}^{+0.21}$ & ${0.52}_{-0.25}^{+0.19}$ & ${1.75}_{-0.16}^{+0.16}$  & $215.40 \pm 75.1 $ &  ${0.37}_{-0.26}^{+0.37}$  & $42.37 \pm 0.11 $ & $41.18 \pm 0.33$ & $42.80 \pm 0.07$ & $42.31\pm0.33$ &x \\
NGC3628 & $64.43/81$ &  $<22.07$ & - & ${2.05}_{-0.09}^{+0.09}$  & $< 100.$ & $< 0.230 $ & $40.18 \pm 0.20 $ & $< 38.4$ & $40.32 \pm 0.27$ & $< 39.3$ &x \\
NGC3718 & $192.45/202$ &  ${23.79}_{-0.13}^{+0.13}$ & $<0.06$ & ${2.21}_{-0.04}^{+0.06}$  & $< 91.1$ & $< 0.111 $ & $41.07 \pm 0.25 $ & $< 39.1$ & $41.43 \pm 0.27$ & $< 40.1$ &x \\
NGC3998 & $613.34/709$ &  ${22.36}_{-0.49}^{+0.74}$ & - & ${1.98}_{-0.04}^{+0.04}$  & $< 18.4$ &  ${0.05}_{-0.03}^{+0.07}$  & $41.59 \pm 0.17 $ & $39.76 \pm 0.27$ & $41.89 \pm 0.16$ & $40.73\pm0.27$ &x \\
NGC4102 & $181.53/194$ &  ${23.68}_{-0.07}^{+0.06}$ & $1*$ & ${1.26}_{-0.11}^{+0.11}$  & $< 312.$ &  ${0.29}_{-0.18}^{+0.22}$  & $41.34 \pm 0.06 $ & $40.00 \pm 0.27$ & $42.08 \pm 0.04$ & $41.50\pm0.27$ &x \\
M106 & $431.61/460$ &  ${23.02}_{-0.03}^{+0.15}$ & $>0.87$ & ${1.86}_{-0.04}^{+0.07}$  & $< 112.$ & $< 0.151 $ & $40.36 \pm 0.03 $ & $< 38.8$ & $40.73 \pm 0.03$ & $< 39.8$ &x \\
M58 & $802.22/732$ &  ${22.34}_{-0.11}^{+0.09}$ & $1*$ & ${2.01}_{-0.04}^{+0.04}$  & $120.73 \pm 21.4 $ &  ${0.08}_{-0.05}^{+0.05}$  & $41.52 \pm 0.01 $ & $39.75 \pm 0.27$ & $41.81 \pm 0.01$ & $40.70\pm0.27$ &x \\
NGC5005 & $26.45/34$ &  ${22.96}_{-0.48}^{+0.39}$ & ${0.68}_{-0.32}^{+0.32}$ & $>2.37$  & $348.49 \pm 145. $ &  $>0.05$  & $40.39 \pm 0.71 $ & $39.58 \pm 0.20$ & $40.50 \pm 0.61$ & $40.25\pm0.20$ &x \\
NGC6240 & $374.82/439$ &  ${24.12}_{-0.03}^{+0.03}$ & ${0.95}_{-0.01}^{+0.01}$ & ${1.63}_{-0.07}^{+0.07}$  & $398.33 \pm 74.3 $ &  ${0.17}_{-0.05}^{+0.06}$  & $43.37 \pm 0.06 $ & $41.87 \pm 0.12$ & $43.87 \pm 0.02$ & $43.08\pm0.12$ &\checkmark \\
MCG+08-31-041 & $200.10/241$ &  $<22.47$ & ${>0.21}$ & ${1.96}_{-0.08}^{+0.09}$  & $150.75 \pm 63.9 $ &  ${0.42}_{-0.20}^{+0.25}$  & $42.64 \pm 0.07 $ & $41.52 \pm 0.20$ & $42.90 \pm 0.04$ & $42.51\pm0.20$ &\checkmark \\
NGC7130\textsuperscript{CT} & $43.65/58$ &  ${24.27}_{-0.25}^{+0.16}$ & $1*$ & $<1.50$  & $715.41 \pm 213. $ &  $>0.42$  & $41.54 \pm 0.38 $ & $40.92 \pm 0.26$ & $42.29 \pm 0.19$ & $42.44\pm0.26$ &\checkmark \\
NGC7331 & $133.31/147$ &  ${22.39}_{-0.43}^{+0.27}$ & $1*$ & ${2.13}_{-0.05}^{+0.16}$  & $< 182.$ & $< 0.437 $ & $40.46 \pm 0.04 $ & $< 39.4$ & $40.69 \pm 0.03$ & $< 40.2$ &x \\
NGC7479\textsuperscript{CT} & $71.56/78$ &  ${24.78}_{-0.06}^{+0.05}$ & $1*$ & ${1.86}_{-0.34}^{+0.61}$  & $693.07 \pm 246. $ &  ${0.20}_{-0.17}^{+0.45}$  & $42.73 \pm 0.48 $ & $40.45 \pm 0.48$ & $42.80 \pm 0.24$ & $41.51\pm0.48$ &\checkmark \\

\hline 

\end{tabular}  	
\caption{Spectral fit results for the LINER sample. Column(1) is the name of the source, (2) is the $\rm{\chi^2/dof}$, where $\rm{dof}$ are the degrees of freedom. Columns (3)-(7) are the column density, $\rm{\log \ NH}$ in units of $\rm{cm^{-2}}$, the covering factor, the photon index, $\Gamma$, the equivalent width (EW) in units of eV and the reflection fraction respectively, while columns (8)-(11) are the continuum and reflected luminosities in units of $\rm{erg \ s^{-1}}$, in the (2-10 keV) and (3-70 keV) bands, respectively. Colum (12) is the f-test performed between models 1 and 3 in all cases. The checkmark (\checkmark) symbol represents a f-test <1E-4, while the x symbol represents otherwise. In Col. (1), the superscript RD is for reflection-dominated sources, while those marked with CT are the Compton-thick sources in our sample, whereas in Col. (4) the dash means that the covering factor is zero, therefore no emission is scattered, and the $*$ symbol means that the parameter is frozen. }
\label{tab:results1}
\end{table*}

\begin{table*}
\centering
\scriptsize
\begin{tabular}{l l l l l l l l l l l l l}\hline 
Name   &  $\rm{\chi^2/dof}$ &   $\rm{\log \ NH}$ & $\rm{f_{cov}}$ & $\rm{\Gamma}$ & $\rm{Eq. \ Width}$ & $\rm{C_{ref}}$ & \multicolumn{2}{c}{$\rm{\log \ L_{(2-10) \ keV}  }$}  & \multicolumn{2}{c}{$\rm{\log \ L_{(3-70) \ keV}  }$} & ftest \\
\cline{8-9} \cline{10-11} \\
& & & & & & & $\rm{continuum}$ & $\rm{reflected}$ & $\rm{continuum}$ & $\rm{reflected}$  \\  

   &   &  $\rm{cm^{-2}}$  &  &   & $\rm{eV}$ & & $\rm{erg \ s^{-1}}$  & $\rm{erg \ s^{-1}}$& $\rm{erg \ s^{-1}}$   & $\rm{erg \ s^{-1}}$  \\
       (1)				&	(2)				&		(3)			&		(4)		&	(5)			&	(6)		&  (7)			&	(8)		&	(9) & (10) & (11) & (12)	 \\ \hline \hline
NGC253 & $566.66/426$ &  ${22.81}_{-0.07}^{+0.04}$ & $>0.84$ & $>2.50$ &  $248.08 \pm 32.8 $ & $< 0.008$ & $39.69 \pm 0.03 $ & $< 36.9$ & $39.76 \pm 0.02$ & $< 37.6$ &x \\
NGC424\textsuperscript{R} & $146.96/139$ &  - & - & ${1.54}_{-0.08}^{+0.08}$ &  $992.62 \pm 155. $ & $>3.52$  & - & $41.31 \pm 0.08$ & - & $42.60\pm0.08$ &\checkmark \\
IC1657 & $224.23/213$ &  ${23.44}_{-0.07}^{+0.06}$ & $1*$ & ${1.52}_{-0.09}^{+0.09}$ &  $234.92 \pm 82.4 $ & ${0.36}_{-0.20}^{+0.24}$  & $41.78 \pm 0.04 $ & $40.58 \pm 0.23$ & $42.35 \pm 0.05$ & $41.88\pm0.23$ &x \\
2MXJ0114-5523 & $133.09/140$ &  ${23.55}_{-0.10}^{+0.09}$ & $1*$ & ${1.26}_{-0.13}^{+0.13}$ &  $< 342.$ & ${0.34}_{-0.23}^{+0.29}$  & $41.73 \pm 0.07 $ & $40.45 \pm 0.30$ & $42.47 \pm 0.05$ & $41.95\pm0.30$ &x \\
MCG+08-03-018 & $181.87/225$ &  ${23.92}_{-0.08}^{+0.07}$ & ${0.90}_{-0.04}^{+0.03}$ & ${2.17}_{-0.13}^{+0.15}$ &  $412.61 \pm 84.1 $ & ${0.37}_{-0.18}^{+0.25}$  & $42.74 \pm 0.12 $ & $41.62 \pm 0.18$ & $42.95 \pm 0.07$ & $42.46\pm0.18$ &\checkmark \\
NGC612 & $138.33/153$ &  ${23.98}_{-0.06}^{+0.05}$ & ${0.99}_{-0.02}^{+0.01}$ & ${1.66}_{-0.13}^{+0.14}$ &  $< 280.$ & $< 0.090$ & $43.36 \pm 0.09 $ & $< 41.5$ & $43.84 \pm 0.03$ & $< 42.7$ &x \\
Mrk573\textsuperscript{CT} & $71.03/60$ &  ${25.04}_{-0.24}^{+0.12}$ & $>0.97$ & ${2.30}_{-0.39}^{+0.14}$ &  $1096.99 \pm 213. $ & ${0.18}_{-0.11}^{+0.26}$  & $42.79 \pm 0.26 $ & $41.39 \pm 0.16$ & $42.95 \pm 0.24$ & $42.18\pm0.16$ &\checkmark \\
NGC788 & $179.07/194$ &  ${23.79}_{-0.08}^{+0.08}$ & $1*$ & ${1.65}_{-0.10}^{+0.11}$ &  $286.80 \pm 97.2 $ & ${0.53}_{-0.23}^{+0.28}$  & $42.51 \pm 0.08 $ & $41.49 \pm 0.16$ & $43.00 \pm 0.04$ & $42.70\pm0.16$ &\checkmark \\
M77\textsuperscript{CT} & $747.04/651$ &  ${25.08}_{-0.03}^{+0.03}$ & ${0.97}_{-0.01}^{+0.01}$ & ${2.37}_{-0.04}^{+0.04}$ &  $1224.43 \pm 64.2 $ & ${0.18}_{-0.03}^{+0.04}$  & $41.93 \pm 0.05 $ & $40.57 \pm 0.04$ & $42.05 \pm 0.03$ & $41.31\pm0.04$ &\checkmark \\
NGC1106\textsuperscript{CT} & $88.06/62$ &  ${24.72}_{-0.11}^{+0.08}$ & $>0.98$ & ${1.64}_{-0.47}^{+0.29}$ &  $1916.85 \pm 423. $ & $>0.11$  & $42.31 \pm 0.68 $ & $41.16 \pm 0.35$ & $42.81 \pm 0.42$ & $42.45\pm0.35$ &\checkmark \\
NGC1125\textsuperscript{CT} & $108.35/119$ &  ${24.42}_{-0.05}^{+0.06}$ & $1*$ & ${2.36}_{-0.39}^{+0.14}$ &  $523.39 \pm 148. $ & ${0.13}_{-0.09}^{+0.16}$  & $42.53 \pm 0.26 $ & $40.99 \pm 0.27$ & $42.63 \pm 0.14$ & $41.73\pm0.27$ &\checkmark \\
NGC1142 & $104.67/115$ &  ${24.07}_{-0.11}^{+0.11}$ & $1*$ & ${1.79}_{-0.16}^{+0.19}$ &  $696.39 \pm 174. $ & ${0.54}_{-0.29}^{+0.42}$  & $42.98 \pm 0.17 $ & $41.98 \pm 0.20$ & $43.38 \pm 0.08$ & $43.07\pm0.20$ &\checkmark \\
Mrk1066 & $45.16/59$ &  ${23.79}_{-0.21}^{+0.19}$ & $1*$ & $<1.42$ &  $< 333.$ & $>0.51$  & $41.27 \pm 0.17 $ & $40.60 \pm 0.26$ & $42.02 \pm 0.13$ & $42.11\pm0.26$ &\checkmark \\
NGC1194 & $148.66/219$ &  ${23.84}_{-0.09}^{+0.09}$ & $1*$ & ${1.53}_{-0.08}^{+0.08}$ &  $631.45 \pm 127. $ & $>0.81$  & $42.06 \pm 0.09 $ & $41.43 \pm 0.12$ & $42.62 \pm 0.07$ & $42.72\pm0.12$ &\checkmark \\
NGC1229 & $129.13/151$ &  ${23.37}_{-0.14}^{+0.12}$ & $1*$ & ${1.42}_{-0.11}^{+0.10}$ &  $448.97 \pm 108. $ & $>0.54$  & $42.58 \pm 0.07 $ & $41.85 \pm 0.18$ & $43.22 \pm 0.07$ & $43.22\pm0.18$ &\checkmark \\
NGC1320\textsuperscript{CT} & $136.01/114$ &  ${24.76}_{-0.17}^{+0.08}$ & $>0.91$ & $>0.35$ &  $3827.63 \pm 345. $ & $>0.35$  & $41.46 \pm 0.53 $ & $40.88 \pm 0.25$ & $42.09 \pm 0.67$ & $42.05\pm0.25$ &\checkmark \\
NGC1358 & $176.34/202$ &  ${24.16}_{-0.06}^{+0.06}$ & $1*$ & ${1.48}_{-0.10}^{+0.11}$ &  $886.71 \pm 165. $ & ${0.47}_{-0.17}^{+0.20}$  & $42.14 \pm 0.10 $ & $41.06 \pm 0.12$ & $42.73 \pm 0.04$ & $42.39\pm0.12$ &\checkmark \\
NGC1386\textsuperscript{CT} & $65.02/39$ &  ${25.15}_{-0.19}^{+0.20}$ & $>0.92$ & ${2.33}_{-0.16}^{+0.14}$ &  $1826.11 \pm 326. $ & ${0.18}_{-0.13}^{+0.58}$  & $41.37 \pm 0.36 $ & $40.01 \pm 0.15$ & $41.41 \pm 0.42$ & $40.76\pm0.15$ &\checkmark \\
UGC3157 & $121.35/158$ &  ${23.57}_{-0.08}^{+0.08}$ & $1*$ & ${1.56}_{-0.12}^{+0.12}$ &  $181.35 \pm 84.4 $ & ${0.31}_{-0.18}^{+0.22}$  & $42.23 \pm 0.07 $ & $40.95 \pm 0.25$ & $42.77 \pm 0.04$ & $42.21\pm0.25$ &x \\
2MSJ0508+1721 & $354.36/363$ &  ${22.49}_{-0.18}^{+0.13}$ & $1*$ & ${1.91}_{-0.07}^{+0.07}$ &  $202.44 \pm 54.3 $ & ${0.40}_{-0.17}^{+0.20}$  & $42.86 \pm 0.02 $ & $41.77 \pm 0.17$ & $43.20 \pm 0.03$ & $42.78\pm0.17$ &\checkmark \\
ESO5-4\textsuperscript{CT} & $128.33/104$ &  $>25.38$ & $>0.98$ & ${1.92}_{-0.09}^{+0.05}$ &  $1622.20 \pm 222. $ & $>0.38$  & $41.53 \pm 0.28 $ & $40.71 \pm 0.07$ & $41.92 \pm 0.27$ & $41.74\pm0.07$ &\checkmark \\
NGC2273\textsuperscript{CT} & $135.57/144$ &  ${25.29}_{-0.14}^{+0.11}$ & $1*$ & ${2.11}_{-0.07}^{+0.05}$ &  $1950.97 \pm 179. $ & ${0.44}_{-0.19}^{+0.37}$  & $42.06 \pm 0.17 $ & $41.01 \pm 0.07$ & $42.26 \pm 0.18$ & $41.90\pm0.07$ &\checkmark \\
UGC3601 & $248.47/252$ &  ${22.61}_{-0.58}^{+0.76}$ & - & ${1.89}_{-0.09}^{+0.08}$ &  $< 211.$ & ${0.16}_{-0.11}^{+0.22}$  & $42.47 \pm 0.86 $ & $41.07 \pm 0.29$ & $42.75 \pm 0.17$ & $42.10\pm0.29$ &x \\
ESO428-14\textsuperscript{CT} & $105.45/107$ &  ${24.48}_{-0.18}^{+0.13}$ & $1*$ & ${2.15}_{-0.24}^{+0.19}$ &  $1100.08 \pm 178. $ & $>0.24$  & $41.41 \pm 0.41 $ & $40.56 \pm 0.19$ & $41.63 \pm 0.31$ & $41.43\pm0.19$ &\checkmark \\
2MXJ0756-4137 & $44.55/58$ &  $<22.55$ & $1*$ & ${1.83}_{-0.10}^{+0.16}$ &  $< 343.$ & $>0.32$  & $41.91 \pm 0.05 $ & $41.07 \pm 0.27$ & $42.30 \pm 0.08$ & $42.16\pm0.27$ &x \\
NGC2788A\textsuperscript{CT} & $72.35/95$ &  ${24.27}_{-0.19}^{+0.14}$ & $1*$ & ${1.41}_{-0.14}^{+0.19}$ &  $1714.57 \pm 303. $ & $>0.55$  & $41.90 \pm 0.28 $ & $41.25 \pm 0.18$ & $42.53 \pm 0.18$ & $42.63\pm0.18$ &\checkmark \\
IC2560\textsuperscript{CT} & $118.75/93$ &  $>25.29$ & $1*$ & ${2.31}_{-0.11}^{+0.04}$ &  $2639.16 \pm 225. $ & ${0.19}_{-0.10}^{+0.29}$  & $42.14 \pm 0.27 $ & $40.72 \pm 0.07$ & $42.39 \pm 0.30$ & $41.51\pm0.07$ &\checkmark \\
NGC3147 & $242.65/288$ &  $<22.65$ & ${0.10}\bullet$ & ${1.84}_{-0.08}^{+0.08}$ &  $126.95 \pm 55.3 $ & $>0.07$  & $41.68 \pm 0.54 $ & $40.31 \pm 0.29$ & $42.15 \pm 0.53$ & $41.38\pm0.29$ &x \\
NGC3393\textsuperscript{CT} & $53.01/83$ &  ${24.37}_{-0.07}^{+0.06}$ & $1*$ & ${1.73}_{-0.19}^{+0.21}$ &  $935.61 \pm 336. $ & ${0.17}_{-0.09}^{+0.12}$  & $42.60 \pm 0.18 $ & $41.07 \pm 0.20$ & $43.04 \pm 0.07$ & $42.21\pm0.20$ &\checkmark \\
2MXJ1105+5856 & $115.59/129$ &  ${23.17}_{-0.52}^{+0.39}$ & ${0.30}\bullet$ & ${1.48}_{-0.15}^{+0.17}$ &  $< 118.$ & $< 1$ & $42.57 \pm 0.30 $ & $< 41.7$ & $43.16 \pm 0.25$ & $< 43.0$ &x \\
NGC3621 & $94.55/94$ &  ${22.73}_{-0.49}^{+0.18}$ & ${0.14}\bullet$ & $>2.46$ &  $< 210.$ & $< 0.219$ & $39.90 \pm 0.32 $ & $< 38.5$ & $40.00 \pm 0.38$ & $< 39.2$ &x \\
NGC3786 & $86.13/98$ &  $<22.45$ & $<0.17$ & ${1.71}_{-0.11}^{+0.11}$ &  $< 161.$ & $< 0.302$ & $42.84 \pm 0.70 $ & $< 40.7$ & $43.42 \pm 0.57$ & $< 41.9$ &x \\
IC751 & $53.50/59$ &  ${23.73}_{-0.10}^{+0.14}$ & $1*$ & $<1.29$ &  $< 733.$ & $>0.25$  & $42.06 \pm 0.12 $ & $41.07 \pm 0.24$ & $42.91 \pm 0.08$ & $42.69\pm0.24$ &x \\
M88\textsuperscript{CT} & $94.56/50$ &  ${24.37}_{-0.20}^{+0.42}$ & ${0.66}_{-0.34}^{+0.25}$ & $>2.44$ &  $< 325.$ & $< 0.344$ & $40.87 \pm 0.55 $ & $< 39.2$ & $40.94 \pm 0.55$ & $< 39.9$ &x \\
IC3639\textsuperscript{CT} & $43.71/46$ &  ${24.80}_{-0.13}^{+0.14}$ & $1*$ & ${2.19}_{-0.23}^{+0.18}$ &  $1896.55 \pm 326. $ & ${0.25}_{-0.15}^{+0.32}$  & $42.03 \pm 0.27 $ & $40.75 \pm 0.19$ & $42.23 \pm 0.19$ & $41.58\pm0.19$ &\checkmark \\
NGC4785 & $171.80/204$ &  ${23.69}_{-0.05}^{+0.05}$ & $1*$ & ${1.79}_{-0.10}^{+0.11}$ &  $187.29 \pm 72.9 $ & ${0.19}_{-0.10}^{+0.12}$  & $42.06 \pm 0.06 $ & $40.61 \pm 0.23$ & $42.46 \pm 0.03$ & $41.71\pm0.23$ &x \\
Mrk231 & $232.65/238$ &  ${22.93}_{-0.11}^{+0.09}$ & $1*$ & ${1.68}_{-0.09}^{+0.09}$ &  $< 228.$ & ${0.28}_{-0.17}^{+0.20}$  & $42.76 \pm 0.03 $ & $41.46 \pm 0.27$ & $43.22 \pm 0.04$ & $42.63\pm0.27$ &x \\
NGC4941 & $111.35/115$ &  ${24.03}_{-0.15}^{+0.14}$ & $1*$ & ${1.65}_{-0.17}^{+0.20}$ &  $688.08 \pm 167. $ & $>0.11$  & $40.81 \pm 0.19 $ & $39.96 \pm 0.23$ & $41.29 \pm 0.10$ & $41.15\pm0.23$ &\checkmark \\
NGC4939 & $158.68/177$ &  ${23.63}_{-0.07}^{+0.06}$ & $1*$ & ${1.51}_{-0.10}^{+0.10}$ &  $204.33 \pm 86.2 $ & ${0.33}_{-0.18}^{+0.22}$  & $41.88 \pm 0.06 $ & $40.63 \pm 0.23$ & $42.45 \pm 0.04$ & $41.94\pm0.23$ &x \\
NGC4945\textsuperscript{CT} & $1477.10/1559$ &  ${24.53}_{-0.01}^{+0.01}$ & ${0.99}\bullet$ & ${1.65}_{-0.03}^{+0.03}$ &  $707.58 \pm 71.7 $ & ${0.05}_{-0.01}^{+0.01}$  & $41.41 \pm 0.02 $ & $39.40 \pm 0.05$ & $41.90 \pm 0.00$ & $40.60\pm0.05$ &\checkmark \\
MCG-03-34-064 & $765.22/722$ &  ${23.88}_{-0.05}^{+0.05}$ & ${0.94}_{-0.01}^{+0.01}$ & ${1.73}_{-0.04}^{+0.05}$ &  $543.42 \pm 44.6 $ & $>0.69$  & $42.74 \pm 0.05 $ & $41.98 \pm 0.05$ & $43.18 \pm 0.03$ & $43.12\pm0.05$ &\checkmark \\
NGC5135\textsuperscript{CT} & $101.51/83$ &  $>25.29$ & $1*$ & ${1.65}_{-0.08}^{+0.09}$ &  $1192.03 \pm 206. $ & $>0.27$  & $40.20 \pm 0.16 $ & $39.21 \pm 0.10$ & $40.77 \pm 0.18$ & $40.41\pm0.10$ &\checkmark \\
M51\textsuperscript{CT} & $85.90/94$ &  ${24.72}_{-0.06}^{+0.13}$ & ${0.81}_{-0.03}^{+0.07}$ & ${1.54}_{-0.06}^{+0.24}$ &  $1636.52 \pm 197. $ & $< 0.047$ & $40.93 \pm 0.10 $ & $37.49 \pm 0.40$ & $41.04 \pm 0.08$ & $37.49\pm1.07$ &\checkmark \\
NGC5252 & $551.16/518$ &  ${23.01}_{-0.43}^{+0.28}$ & ${0.38}\bullet$ & ${1.77}_{-0.05}^{+0.07}$ &  $< 114.$ & $< 0.187$ & $43.17 \pm 0.15 $ & $< 41.6$ & $43.58 \pm 0.16$ & $< 42.7$ &x \\
NGC5283 & $216.31/230$ &  ${23.10}_{-0.07}^{+0.07}$ & $1*$ & ${1.92}_{-0.08}^{+0.08}$ &  $113.85 \pm 67.7 $ & ${0.42}_{-0.20}^{+0.24}$  & $41.98 \pm 0.03 $ & $40.90 \pm 0.19$ & $42.31 \pm 0.04$ & $41.91\pm0.19$ &\checkmark \\
NGC5347\textsuperscript{CT} & $43.70/45$ &  ${24.79}_{-0.13}^{+0.15}$ & $>0.94$ & ${1.74}_{-0.13}^{+0.13}$ &  $2299.24 \pm 326. $ & ${0.09}_{-0.06}^{+0.12}$  & $42.03 \pm 0.27 $ & $40.06 \pm 0.26$ & $42.23 \pm 0.19$ & $41.11\pm0.26$ &\checkmark \\
NGC5643\textsuperscript{CT} & $156.54/142$ &  ${24.94}_{-0.23}^{+0.25}$ & $1*$ & ${2.47}_{-0.09}^{+0.06}$ &  $1137.79 \pm 151. $ & ${0.28}_{-0.12}^{+0.24}$  & $41.35 \pm 0.21 $ & $40.24 \pm 0.07$ & $41.42 \pm 0.19$ & $40.93\pm0.07$ &\checkmark \\
NGC5695 & $38.99/38$ &  ${23.94}_{-0.11}^{+0.39}$ & $>0.77$ & $<1.40$ &  $797.16 \pm 644. $ & $>0.75$  & $41.05 \pm 0.31 $ & $40.45 \pm 0.26$ & $41.87 \pm 0.21$ & $42.07\pm0.26$ &x \\
NGC5728 & $336.17/398$ &  ${24.05}_{-0.03}^{+0.03}$ & $1*$ & ${1.45}_{-0.06}^{+0.06}$ &  $428.24 \pm 83.7 $ & ${0.26}_{-0.07}^{+0.07}$  & $42.21 \pm 0.05 $ & $40.87 \pm 0.09$ & $42.83 \pm 0.02$ & $42.23\pm0.09$ &\checkmark \\
NGC5899 & $308.81/378$ &  ${22.98}_{-0.16}^{+0.25}$ & ${0.87}_{-0.32}^{+0.13}$ & ${1.80}_{-0.07}^{+0.07}$ &  $95.88 \pm 46.9 $ & ${0.28}_{-0.14}^{+0.22}$  & $42.16 \pm 0.08 $ & $40.90 \pm 0.17$ & $42.56 \pm 0.10$ & $41.99\pm0.17$ &\checkmark \\
MCG+14-08-004 & $130.22/161$ &  ${23.24}_{-0.11}^{+0.10}$ & $1*$ & ${1.87}_{-0.12}^{+0.12}$ &  $265.59 \pm 81.6 $ & ${0.31}_{-0.19}^{+0.24}$  & $42.48 \pm 0.06 $ & $41.25 \pm 0.27$ & $42.84 \pm 0.05$ & $42.28\pm0.27$ &x \\
ESO137-34\textsuperscript{CT} & $96.36/106$ &  ${24.49}_{-0.09}^{+0.12}$ & $>0.98$ & ${1.78}_{-0.22}^{+0.20}$ &  $1066.66 \pm 216. $ & ${0.31}_{-0.18}^{+0.32}$  & $42.14 \pm 0.26 $ & $40.92 \pm 0.19$ & $42.54 \pm 0.15$ & $42.02\pm0.19$ &\checkmark \\
2MXJ1650+0436 & $331.00/392$ &  $<22.45$ & ${0.52}\bullet$ & ${1.72}_{-0.06}^{+0.06}$ &  $93.32 \pm 46.1 $ & ${0.24}_{-0.12}^{+0.14}$  & $43.29 \pm 0.04 $ & $41.95 \pm 0.21$ & $43.73 \pm 0.02$ & $43.09\pm0.21$ &x \\
2MXIJ1802-1454 & $605.51/591$ &  ${22.45}_{-0.11}^{+0.09}$ & $1*$ & ${2.02}_{-0.05}^{+0.05}$ &  $97.45 \pm 32.1 $ & ${0.18}_{-0.07}^{+0.08}$  & $41.76 \pm 0.01 $ & $40.35 \pm 0.16$ & $42.04 \pm 0.02$ & $41.29\pm0.16$ &\checkmark \\
2MXJ1830+0928 & $158.71/179$ &  ${23.18}_{-0.07}^{+0.07}$ & $1*$ & ${1.89}_{-0.10}^{+0.10}$ &  $224.05 \pm 73.6 $ & ${0.35}_{-0.19}^{+0.24}$  & $42.37 \pm 0.04 $ & $41.21 \pm 0.23$ & $42.72 \pm 0.04$ & $42.23\pm0.23$ &x \\
IC4995\textsuperscript{CT} & $66.10/65$ &  $>25.30$ & ${0.97}_{-0.02}^{+0.01}$ & ${1.81}_{-0.07}^{+0.14}$ &  $1266.81 \pm 251. $ & $>0.14$  & $42.43 \pm 0.29 $ & $41.26 \pm 0.20$ & $42.83 \pm 0.26$ & $42.34\pm0.20$ &\checkmark \\
NGC6921 & $113.83/113$ &  ${24.16}_{-0.09}^{+0.08}$ & $1*$ & ${1.66}_{-0.16}^{+0.19}$ &  $692.22 \pm 187. $ & ${0.34}_{-0.17}^{+0.21}$  & $42.34 \pm 0.15 $ & $41.13 \pm 0.20$ & $42.82 \pm 0.04$ & $42.32\pm0.20$ &\checkmark \\
NGC7213 & $879.92/959$ &  ${22.57}_{-0.45}^{+0.10}$ & - & ${1.97}_{-0.03}^{+0.03}$ &  $80.68 \pm 15.7 $ & ${0.04}_{-0.02}^{+0.04}$  & $42.00 \pm 0.19 $ & $40.14 \pm 0.17$ & $42.31 \pm 0.19$ & $41.11\pm0.17$ &x \\
NGC7319 & $74.48/83$ &  ${23.62}_{-0.06}^{+0.22}$ & $>0.83$ & ${1.38}_{-0.12}^{+0.16}$ &  $710.94 \pm 159. $ & $>0.48$  & $41.40 \pm 0.11 $ & $40.61 \pm 0.20$ & $42.02 \pm 0.09$ & $42.00\pm0.20$ &\checkmark \\
UGC12282 & $38.97/39$ &  ${24.15}_{-0.25}^{+0.16}$ & $1*$ & ${1.62}_{-0.22}^{+0.30}$ &  $1602.02 \pm 441. $ & $>0.35$  & $41.70 \pm 0.38 $ & $41.03 \pm 0.29$ & $42.20 \pm 0.25$ & $42.25\pm0.29$ &\checkmark \\
NGC7582 & $993.71/1095$ &  ${23.48}_{-0.03}^{+0.03}$ & ${0.96}_{-0.01}^{+0.01}$ & ${1.60}_{-0.03}^{+0.03}$ &  $83.17 \pm 21.9 $ & ${0.19}_{-0.04}^{+0.04}$  & $42.16 \pm 0.01 $ & $40.70 \pm 0.08$ & $42.68 \pm 0.01$ & $41.94\pm0.08$ &\checkmark \\
2MXJ2325-3826 & $62.91/76$ &  $<22.24$ & $1*$ & ${1.93}_{-0.11}^{+0.13}$ &  $< 219.$ & $< 1$ & $42.13 \pm 0.03 $ & $< 41.4$ & $42.46 \pm 0.07$ & $< 42.4$ &x \\
NGC7674 & $154.67/156$ &  ${23.05}_{-0.14}^{+0.12}$ & $1*$ & ${1.57}_{-0.09}^{+0.09}$ &  $258.83 \pm 90.5 $ & $>0.75$  & $41.80 \pm 0.05 $ & $41.19 \pm 0.15$ & $42.33 \pm 0.07$ & $42.45\pm0.15$ &\checkmark \\
\hline 
\end{tabular}  	
\caption{Spectral fit results for the Seyfert sample. Columns are as in Tab.\,\ref{tab:results1}.}
\label{tab:results2}
\end{table*}

Tables\,\ref{tab:results1} and \ref{tab:results2} present the best-fit parameters for both the LINER and Seyfert samples when fitted to the $\rm{M_3}$ model. Note that we present the best-fit values for this model only as the analysis is based on this model version (see above).  Note that for the covering factor, $\rm{f_{cov}}$, we also test whether is zero or free to vary. In the cases in which it is zero, the source is not covered, corresponding to unobscured/low-obscured sources or, on the contrary, it is reflection dominated, thus the partially-covered continuum is not restricted. We mark the column with a dash line in these cases (six sources). However, although left free to vary, in many cases it is pegged to one (42 sources), indicating full covering of the source. 
As for the f-test, we conclude that when it is below $\rm{10^{-3}}$, the reflection component is statistically significant and is marked with the check mark symbol in this Col.(12). Among the 81 objects, we find that only NGC\,424 is reflection dominated, i.e., the data do not require the addition of a power-law and the luminosity of this component is not restricted. This object is marked with an R in the Col.\,(1) of Table\, \ref{tab:results2}. The $\rm{N_H}$ cannot be calculated for this source. Moreover, we also find 18 other sources for which the f-test comparing models $\rm{M_3}$ and $\rm{M_4}$ does not favour the addition of a power-law component, although their $\rm{N_H}$ are well constrained. These objects have a large contribution of the reflection component in the CT regime, although the power-law continuum is still reachable thanks to the wide spectral range of \textit{NuSTAR} data. {As for intrinsic luminosities, as it was previously mentioned in section\,\ref{sec:spectral-fitting}, these may be larger for CT objects as we are inducing a systematic uncertainty by excluding the Compton-scattering effect. Thus, we warn the reader that intrinsic luminosities reported here for CT objects should be treated a lower limits, even if they are well constrained by the spectral fits.}

\subsection{Luminosity Correction}\label{sec:luminositycorrection}

\begin{figure}
    \centering
    \includegraphics[width = 1.\columnwidth]{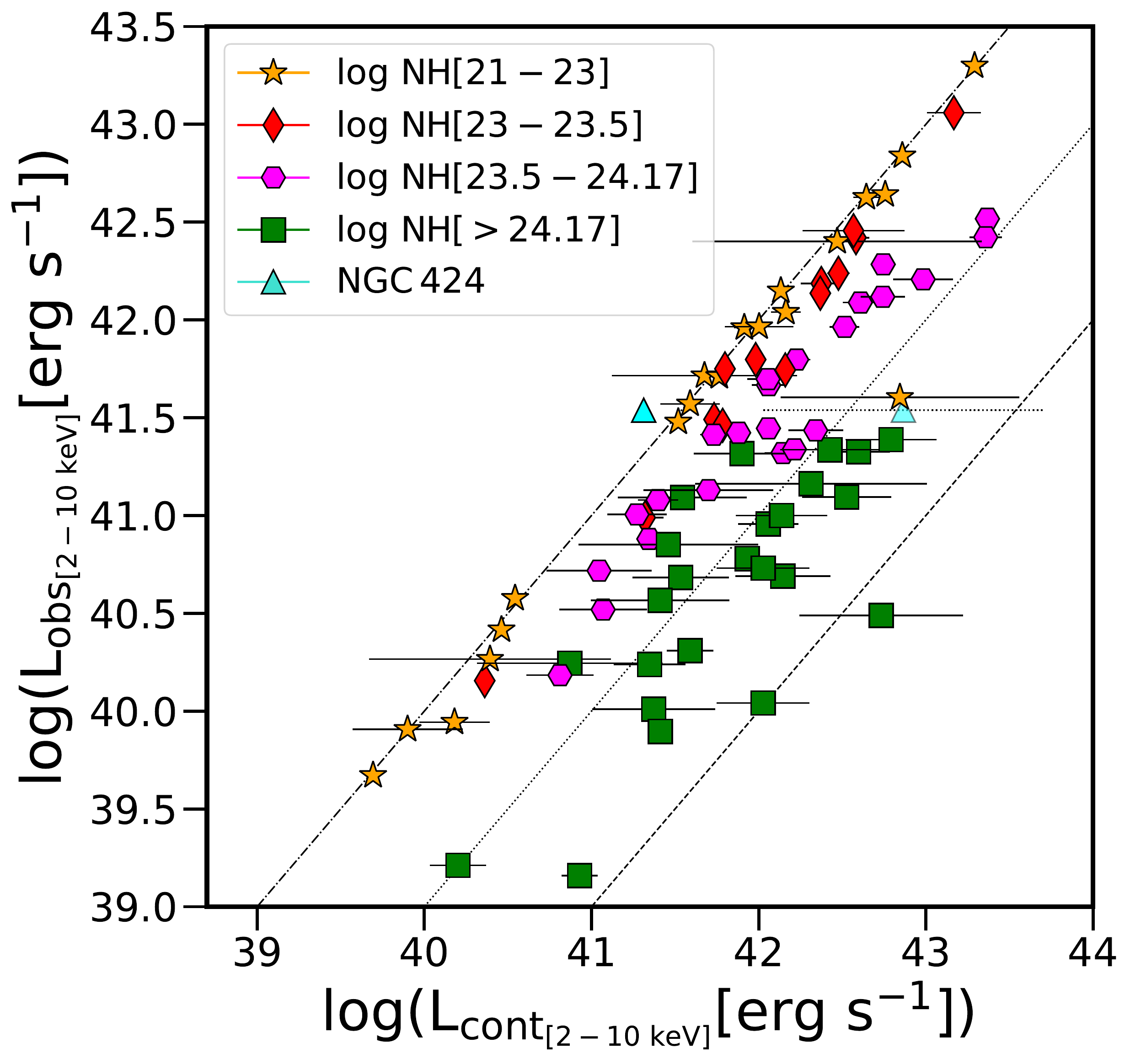}
    \caption{Observed 2-10\,keV luminosity versus the intrinsic 2-10\,keV luminosity. Note that the upper limits refer to the reflection dominated objects (see text). The dash-dotted, dotted, and dashed lines represent the 1:1, 1:10, 1:100 relations, respectively, indicating the correction factor that should be applied in order to obtain the intrinsic continuum luminosity from the observed luminosity. Orange stars, red diamonds, magenta hexagons and green squares are the $\rm{N_H}$ ranges shown in the figure label, while the cyan triangles are NGC\,424 uncorrected (solid) and corrected (semi-translucent, see text).} 
    \label{fig:luminosity-comparison}
\end{figure}

As mentioned in Section\,\ref{sec:sample} we select sources with Eddington rate $\rm{L_{bol,obs}/L_{Edd}<10^{-3}}$ to study inefficient AGN with the public data available. However, we use as a proxy of the bolometric luminosity, the observed 2-10\,keV X-ray luminosity (i.e. uncorrected from $\rm{N_H}$). Indeed, the intrinsic luminosity and therefore the Eddington rate are expected to increase once we take into account the $\rm{N_H}$. Fig.\,\ref{fig:luminosity-comparison} shows the observed versus intrinsic 2-10\,keV luminosity, the latter corrected from line-of-sight (LOS) obscuration according to the best fit reported in Tables \ref{tab:results1} and \ref{tab:results2}. 

The largest shift from the 1:1 relation is found for CT objects (green squares in Fig.\,\ref{fig:luminosity-comparison}), for which the luminosity can change more than a factor of ten. Note that NGC\,424, classified as reflection dominated has long been studied at X-rays \citep[e.g., ][]{Balokovic-14, Hernandez-15, Ricci-17, Marchesi-18}, finding that this object has a large contribution of the reflection component, with weak signatures of the intrinsic continuum.  

However, since we are unable to account for such a component, we calculate the bolometric luminosity through the luminosity of the reflection component as follows: we assume that the correction in the intrinsic luminosity for a certain object is given in the following form: $\rm{\log(L_{cont}) =  A+ \log(L_{refl})}$, where $\rm{A}$ can be calculated through those CT objects for which the power-law is still detected in the spectra.

Moreover, by calculating the constant $\rm{A}$ for these objects, we can estimate how much the intrinsic luminosity changes for a certain amount of reflection. In practice, we estimate this factor using the green objects in Fig.\,\ref{fig:luminosity-comparison} (i.e. 25 CT AGN). We find for our sample that $\rm{A = 1.5 \pm 0.8}$. We then use this value to estimate the intrinsic luminosity for NGC\,424. This extrapolation can be seen as the semi-transluscent cyan triangle in Fig.\,\ref{fig:luminosity-comparison}. 

\begin{figure}
    \centering
    \includegraphics[width = 0.96\columnwidth]{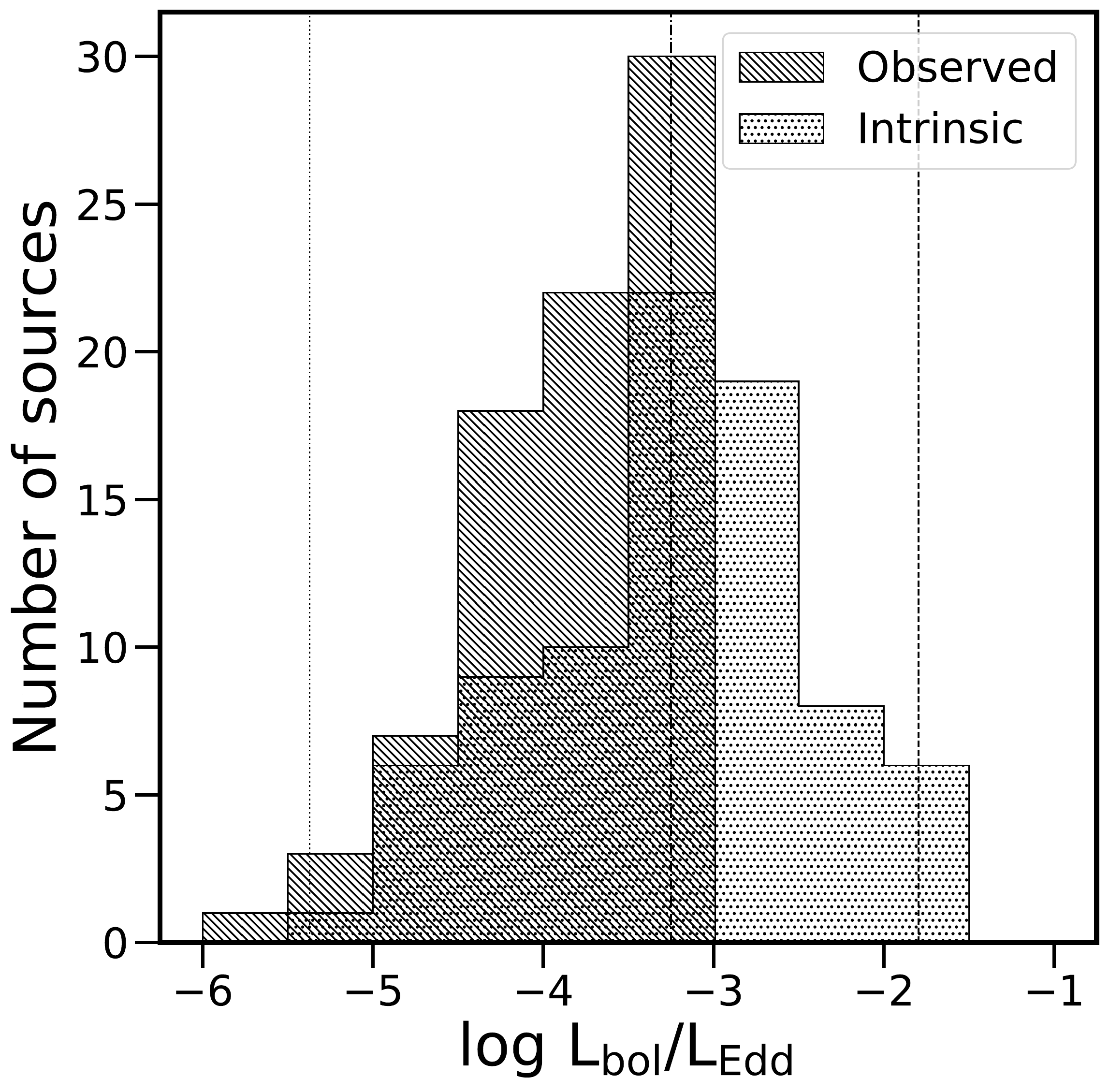}
    \caption{Comparison between the Eddington rate obtained with a rough fit to a simple power-law (dashed histogram) versus the Eddington rate obtained once a proper obscuration is accounted for (dotted histogram). }
    \label{fig:Eddington-comparison}
\end{figure}

Fig.\,\ref{fig:Eddington-comparison} shows the final  distribution for the Eddington rate in our sample when the obscuration is properly taken into account, compared to the initial Eddington rate, estimated through the observed X-ray luminosity. 
Indeed, a shift towards greater Eddington rates is seen in objects with larger $\rm{N_H}$ (mostly CT objects) and a decrease in objects with lower Eddington rates. The mean Eddington rate of the sample is $\rm{<log \, (\lambda_{Edd})> = -3.3}$ (dash-dotted vertical line in Fig.\,\ref{fig:Eddington-comparison}) with a 1$\sigma$ range of $[-5.2, -1.49]$ (see dotted lines in the figure), roughly expanding four orders of magnitude in Eddington rate. {Additionally, the lack of Compton-scattering may shift CT objects to even higher Eddington rates (as intrinsic luminosities may be even larger)}. As it was previously mentioned in section \ref{sec:data-red}, we are aware that our selection criteria bias our sample towards a large LOS obscuration for Eddington rate between $\rm{-2.5  < \log \lambda_{Edd} < -1.49}$ (14 CT AGN). Thus, we will exclude these sources from the discussion as we are missing an important portion of objects residing in this range of Eddington rates (mostly highly obscured objects with intrinsically higher Eddington rates). {Moreover, the discussion will be based on the remaining 67 sources.}

\begin{figure*}
\includegraphics[width = 1\linewidth]{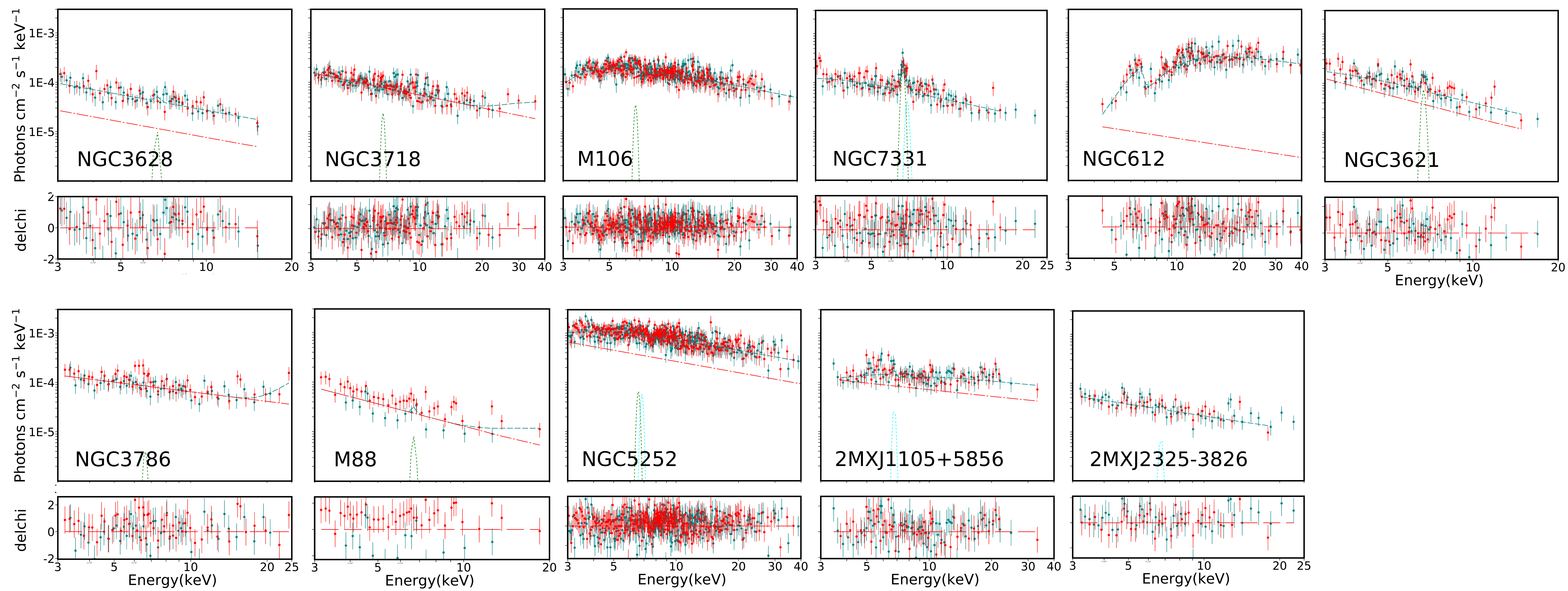}
      \caption{Spectral fits for our eleven candidates for the disappearance of the torus.  
      Colors and axis labels are as in Fig.\,\ref{fig:models}. Note that we have ionized lines are present in all the objects (except for NGC\,612) and that we have tested the scenario in which the FeK$\alpha$ line could be the responsible for the line feature, finding that such line is not present in any of the spectra. Also note that for M\,88 both FPMA and FPMB modules seem to be misaligned with a normalization factor of 1.99, well above the overall sample.} 
  \label{fig:candidates}
\end{figure*}

\subsection{Existence of the reflection component}\label{sec:existence}

We base our analysis on the existence of the reflection component in three criteria: (1) the detection of the FeK$\rm{\alpha}$ line; (2) a statistical significance of the reflection component compared to the model without reflection (i.e. f-test $<0.001$ when comparing models $\rm{M_1}$ and $\rm{M_3}$ ), and (3) the restriction of the reflection luminosity at 3--$\rm{\sigma}$ .  

We detect the FeK$\rm{\alpha}$ emission line in 61 out of the 81 sources ($\rm{\sim 75\%}$ of the sample). On the other hand, the inclusion of a reflection component improves the spectral fit for 46 out of the 81 sources ($\sim 57\%$ of the sample). Additionally, we restrict the luminosity of the reflection component for 67 out of the 81 sources (i.e. $\rm{\sim 83\%}$). A total of 45 out of 81 ($\rm{\sim 55\%}$) sources fulfill the three criteria to establish the existence of the reflection component, while 15 ($\rm{\sim 19\%}$) meet two out of the three criteria, and 9 ($\rm{\sim 10\%}$) meet only one of the criteria (namely M\,81, NGC\,3998, NGC\,4102, NGC\,253, 2MXJ\,0114-5523, UGC\,3601, 2MXJ\,0756-4137, IC\,751, and Mrk\,231). 

As for the 15 sources fulfilling two criteria, 
one (Mrk\,1066) statistically needs the reflection component and it is well constrained. However, the $\rm{FeK\alpha}$ emission line is not well detected at a $3-\sigma$ limit. Note that the spectrum is a low-quality spectrum (dof$\rm{<}$50) which may cause a poor restriction of the three lines at 6.4, 6.7 and 6.97 keV. In order to test this possibility, we use the non-grouped spectrum of the source and check which line is present. Indeed, the line present in this object is the $\rm{FeK\alpha}$ emission line. Once we remove both ionized lines from the fit, and calculate the EW and flux for the $\rm{FeK\alpha}$ emission line for this source, we obtain a constrained parameter. Thus, we can safely say that this source fulfills the three criteria.

The remaining 14 objects, with two out of the three criteria, fulfill the detection of the $\rm{FeK\alpha}$ emission line and a constrained luminosity for the reflection component. However, statistically speaking, the reflection component is not needed. Among them, all but NGC\,5695 show a relatively low reflection fraction ($\rm{C_{ref}}$) and/or $\rm{N_H}$. Thus, the reflection component is present mainly through the $\rm{FeK\alpha}$ emission line while the Compton hump is suppressed under the intrinsic continuum. Under this scenario, the f-statistic is not suitable to test the presence of the reflection component because it only affects a narrow range of the spectrum where the $\rm{FeK\alpha}$ emission line is located. In the case of NGC\,5695, it shows a relatively high $\rm{N_H}$. In this case, the quality of the spectrum for this source might be responsible for the statistical significance of the reflection component. 

As for the objects fulfilling only one criterion, seven objects (NGC\,3998, NGC\,4102, 2MXJ\,0114-5523, UGC\,3601, 2MXJ\,0756-4137, IC\,751 and Mrk\,231) meet the restriction of the luminosity of the reflection component, and the remaining two objects (M\,81 and NGC\,253) meet the detection of the $\rm{FeK\alpha}$ emission line. The detection of the reflection component without the $\rm{FeK\alpha}$ line can be due to a combination of high absorption which can lead to the detection of a Compton hump and a low S/N spectrum which can lead to the poor restriction of the line. The detection of the $\rm{FeK\alpha}$ line without statistical significance for the reflection component can again be explained due to low $\rm{N_H}$ and/or low reflection fraction as indicated above. Moreover, the non-detection of the Compton hump may also depend on the shape of the spectrum; for highly obscured sources it may only require data up to 10 keV, while for low obscured sources, the Compton hump requires high S/N up to 30 keV. 

We find that 12 objects do not present any signs of reflection (namely UGC\,5101, NGC\,3628, NGC\,3718, M\,106, NGC\,7331, NGC\,612, 2MXJ\,1105+5856, NGC\,3621, NGC\,3786, M\,88, NGC\,5252 and 2MXJ\,2325-3826). Note that five out these 12 objects are LINERs while the remaining seven are Seyferts. However, several works have found traces for the reflection component in UGC\,5101 \citep[e.g.,][]{Gonzalez-09, Oda-17, LaCaria-19}. In addition, the fact that the spectrum has a low S/N leads us to exclude it from our final sample of candidates for the disappearance of the torus. Fig.\,\ref{fig:candidates} shows the spectral fits for our final sample of 11 candidates to the torus disappearance  ($\rm{\sim 12\%}$). All but NGC\,612 can be modelled with a single power-law\footnote{Note that all of them present ionized emission lines.} with little obscuration, supporting the idea that the reflection component is not present in these objects.

Altogether, we present 11 candidates which lack the reflection component, related to the plausible disappearance of the torus using X-rays observations. Among them, we find that eight sources have studies aiming to find the reflection component or the torus (through X-rays/mid-IR). For all of them we find complete agreement with the idea of the absence of the reflection component associated with the torus in previous works (see Sec.\,\ref{sec:discussion}).

\begin{figure*}
\centering
\includegraphics[width=0.8\linewidth]{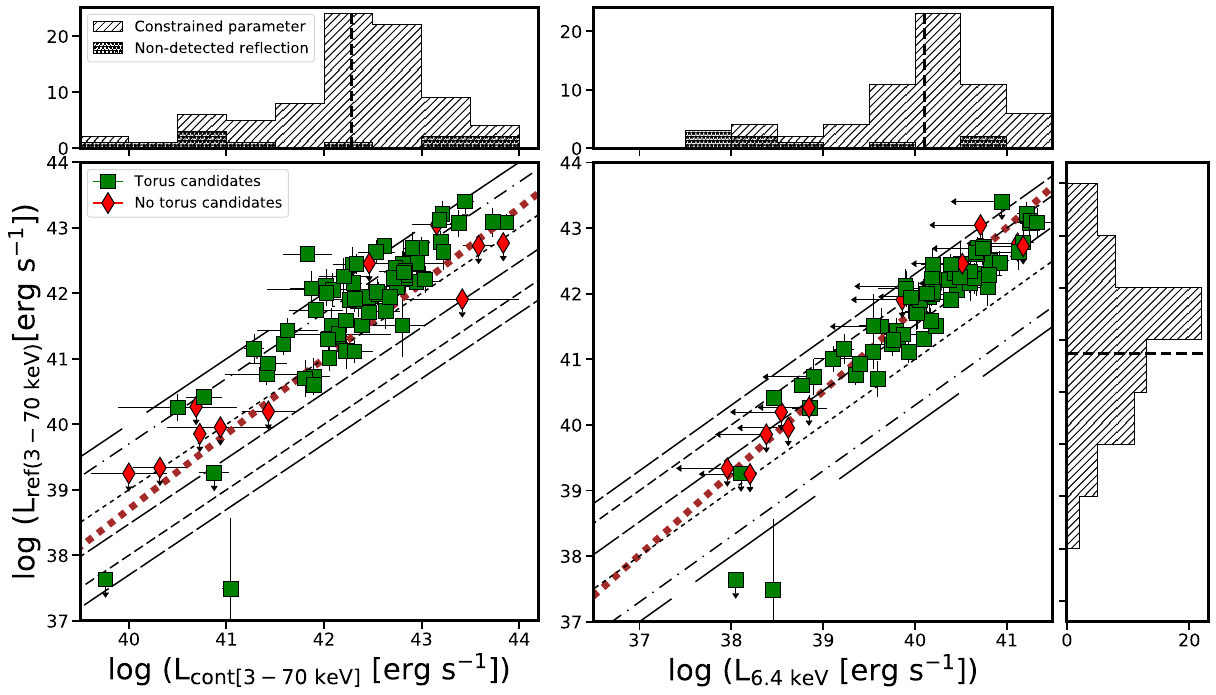}
   
    \caption{3-70 keV band luminosity of the reflection component versus that of the intrinsic continuum (left) and the $\rm{FeK\alpha}$ emission line luminosity versus the 3-70 keV  reflection luminosity (right). In both cases, lines represent the percentage of the quantity in the Y axis with respect to the X axis. From top to bottom 100, 50, 10, 3, 1, and 0.5\% of the intrinsic (large left panel) and reflection (large right panel) luminosities. The red diamond and green square are objects well fitted with the power-law components (i.e. no reflection candidates) and with the combination of reflection plus power-law components, respectively. The thick dotted brown line in both panels represent the linear correlation found for the quantities in each panel when the X axis is the independent quantity, while the black dashed line in the histograms represents the mean of each quantity. The histograms in the top and right panels are the distribution of the different luminosities (see text).}
    \label{fig:reflvslfe}
  \end{figure*}
  \subsection{Reflection strength}
We explore here the degree of the reflection component strength for the objects in our sample. { Fig.\,\ref{fig:reflvslfe} shows the luminosity distribution for the intrinsic continuum (left histogram), $\rm{FeK\alpha}$ emission line (upper right histogram) and reflection (right histogram). Note that for both the reflection and FeK$\alpha$ line distributions, we only show those sources for which the parameter is well constrained.} 

The $\rm{FeK\alpha}$ line luminosity ranges from $\rm{log(L_{6.4 \ keV})\sim [38.-41.3]}$, with a mean value of $\rm{<log(L_{6.4 \ keV})>= 40.1}$, the reflection luminosity ranges from $\rm{log(L_{ref(3-70 \ keV)})\sim [37.5,43.2]}$, with a mean value of $\rm{<log(L_{ref(3-70 \ keV)})> = 41.9}$, and the intrinsic 3-70\,keV luminosity ranges from $\rm{log(L_{cont(3-70 \ keV)})\sim [39.8,43.9]}$, with a mean value of $\rm{<log(L_{cont(3-70 \ keV)})> = 42.3}$ (see dashed lines in the histograms). Thus, all these three luminosities span around four orders of magnitude. 

 The reflection component represents, on average, 40\% of the intrinsic continuum luminosity of the source, (i.e. $\rm{<C_{ref}> = 0.41}$\footnote{Note that this value is calculated for objects with both luminosities well constrained (i.e., neglecting upper/lower limits.)}). In 20 out of the 81 sources the reflection is below 20\% of the intrinsic continuum (i.e. $\rm{C_{ref} < 0.2}$), 28 sources fall in the range $\rm{0.2 < C_{ref} < 0.6}$, and 5 fall in the range $\rm{C_{ref} > 0.6}$, while the remaining 28 sources have lower limits for $\rm{C_{ref}}$.
On the other hand, around 50\% of the sample (41 sources) is located in the area where FeK$\rm{\alpha}$ luminosity accounts for 1-3\% of the reflection component luminosity, 15\% (12 sources) is consistent with 3-10\%, and 6\% (five sources) sources present a weak line luminosity, accounting for less than 1\% of the reflection component. As for the remaining 28\% (23 sources), either the EW of the FeK$\rm{\alpha}$ line and/or the luminosity of the reflection component are upper limits, therefore the fraction is an upper limit as well. Naturally, the 11 candidates lacking of reflection signatures are included in this group. 

 In order to study the connection between these luminosities, the large panels in Fig.\,\ref{fig:reflvslfe} show the correlation between the reflection luminosity versus the intrinsic continuum (left panel) and the FeK$\rm{\alpha}$ emission line (right panel) luminosities. As expected, the three quantities are well correlated. Moreover, the reflection component and the FeK$\rm{\alpha}$ emission line are slightly better correlated (with a Pearson's coefficient of 0.93) than the intrinsic continuum and the reflection component (with a Pearson's coefficient of 0.89). 

\noindent  In order to obtain the best characterization of the relation between those parameters, we use a bootstrap method to account for possible outliers in the correlations, by choosing a random subsample with 80\% of the total sample and repeating this process a total of 100 times. For each random subsample, we also perform a MCMC simulation similar to that mentioned in section \ref{sec:spectral-fitting}: for each luminosity pair, we choose a random number drawn from a normal distribution whenever the value is well constrained, and from an uniform distribution whenever there are upper/lower limits involved. We then perform a binning method in the {\it x} axis by dividing it into bins of $dx = 0.5$ dex, and obtain the median and standard deviation along both axes. We calculate the correlation through a linear regression method and obtain the slope, intercept, and the r- and p-values. We repeat this process 100 times and then estimate the mean values of all quantities. As for the correlation between $\rm{L_{ref}}$ and $\rm{L_{6.4 \ keV}}$ we also perform the linear regression analysis assuming that the luminosity of the reflection component is the independent component and the luminosity of the line is the dependent one. We present the best correlation found after performing a bisection method from both correlations. In this case the fraction of the luminosity of the FeK$\rm{\alpha}$ line over the Compton hump increases from 1\% at $\rm{log(L_{ref})=43}$ to $\rm{\sim}$5\% at $\rm{log(L_{ref})=40}$ and to $\rm{\sim}$10\% at $\rm{log(L_{ref})=38}$. Both correlations can be represented in the following form (represented as the thick dotted brown lines in Fig.\,\ref{fig:reflvslfe}):
\begin{multline}
\rm{\log(L_{ref}) = (1.13\pm0.22)\log(L_{cont}) - (6.43\pm9.35)} 
\end{multline}
\begin{multline}
\rm{\log(L_{ref}) = (1.24\pm0.15)\log(L_{6.4 \ keV}) - (8.17\pm6.16)}
\end{multline}

Note that we also explore if this trend is seen over the intrinsic continuum or the Eddington rate but no correlation is found.

\subsection{Obscuration}

\begin{figure}
    \includegraphics[width = 0.99\columnwidth]{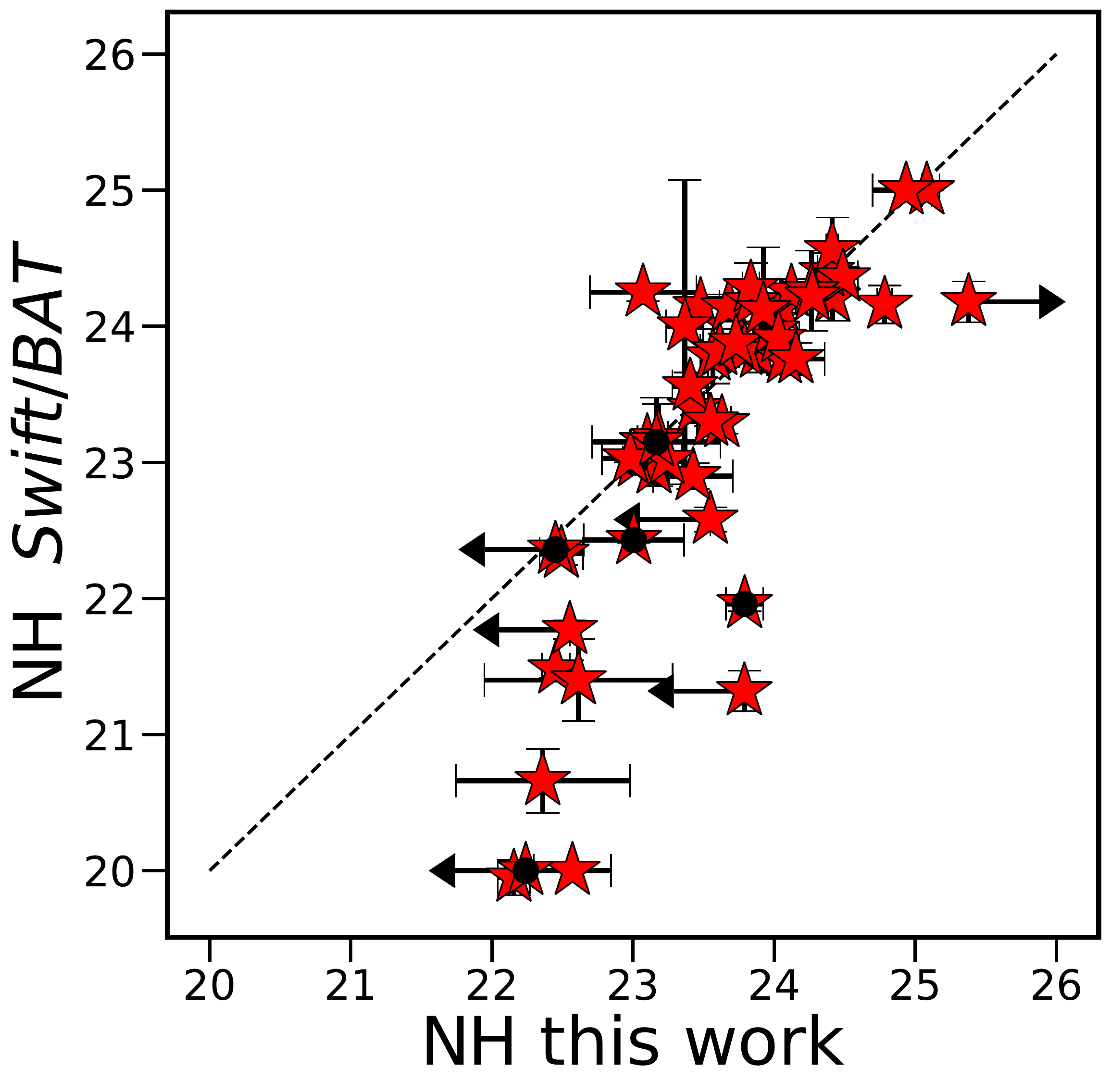}
    \caption{Comparison between our $\rm{N_H}$ estimates and the values reported in \citet{Ricci-17} for the common sample. Sources with a black dot inside the red star {are the } candidates for the torus disappearance in our sample. Note that upper limits in our analysis might be due to the fact that the sources are low-obscured or unobscured (see text).} 
    \label{fig:nhcomparison}
\end{figure}

\begin{figure}
\centering
    \includegraphics[width=1.0\columnwidth]{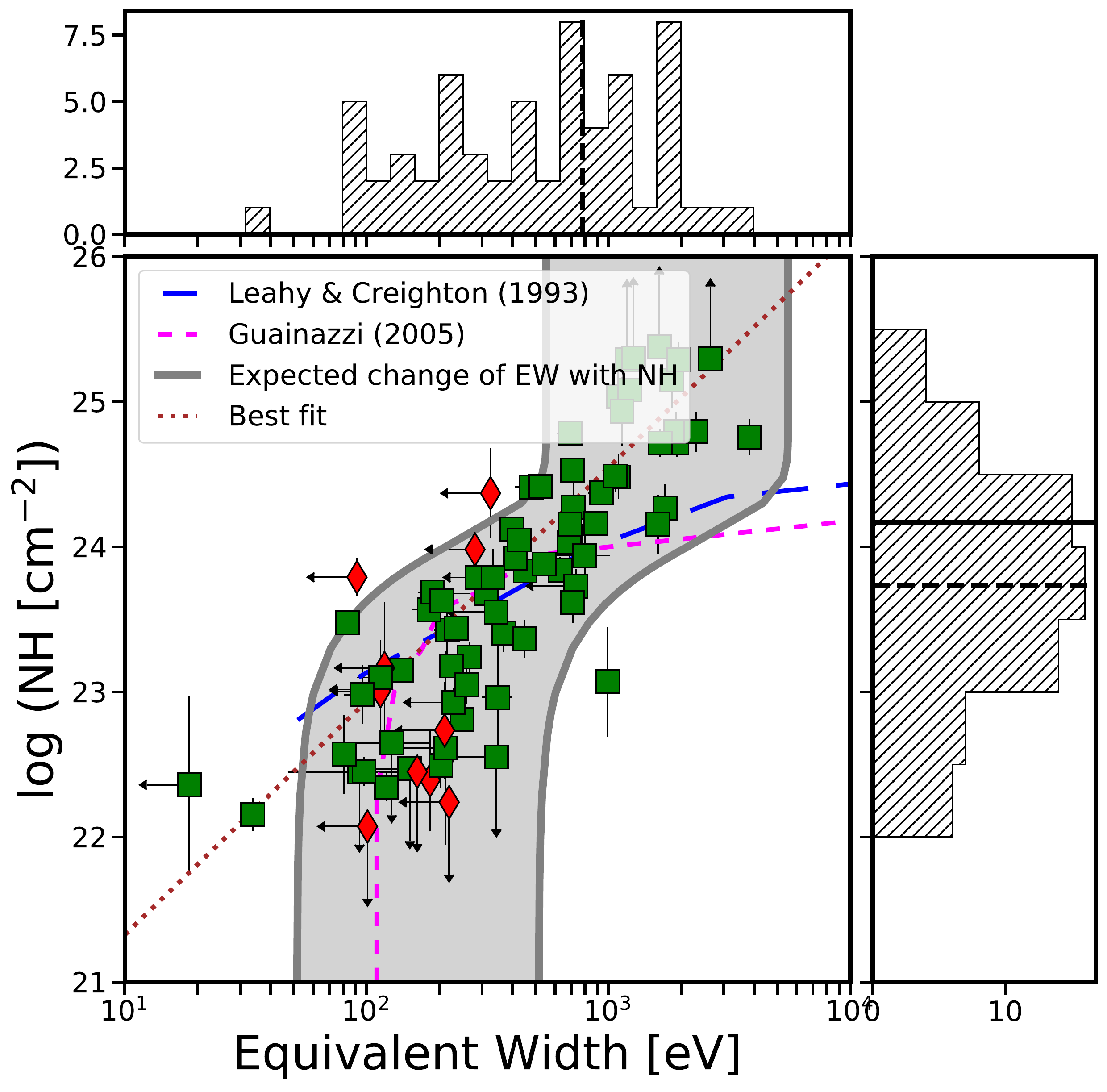}

    \caption{Equivalent width of the $\rm{FeK\alpha}$ line versus the column density along the LOS. Colors and symbols are as in Fig.\,\ref{fig:reflvslfe}. The correlation line is drawn as the brown dotted line while the blue long-dashed line and pink dashed lines show the expected relation according to \citet{Leahy93} and \citet{Guainazzi-05}, respectively. Moreover, the gray shaded area corresponds to the case in which the EW changes according to different levels of obscuration and reflection (see text).}
    
     \label{fig:nhvseq}
  \end{figure}

The LOS obscuration and the reflection might occur in the same structure (i.e., the torus), thus both the reflection and the obscuration may be related. Therefore, a proper estimate on the $\rm{N_H}$ is necessary in order to place solid conclusions on the reflection component. For this purpose, Fig.\,\ref{fig:nhcomparison} shows the comparison between the $\rm{N_H}$ found in this analysis compared to those presented in \citet{Ricci-17}. They analyze the X-ray properties of the \textit{Swift}/BAT sample for which we find 53 sources in common. We find that for most sources, the LOS obscuration found in both works are well in agreement when their column density is $\rm{N_H>3\times10^{22} cm^{-2}}$. We find discrepancies  for seven objects below $\rm{N_H<3\times10^{22} cm^{-2}}$, which may be expected since \emph{NuSTAR} does not cover energies below 3 keV, necessary to constrain low values of obscuration. 
Moreover, Fig.\,\ref{fig:nhvseq} (right histogram) shows the $\rm{N_H}$ distribution for our sample. We find a mean column density of $\rm{< log(N_H)> = 23.74}$ (see dashed line in the figure). We also find that 24 sources are CT AGN (i.e., $\rm{\log N_H > 24.17 \ cm^{-2}}$) while 45 out of the 81 are Compton-thin and 12 are unobscured sources. 
In addition, the majority of sources with no clear indication of reflection appear to be below the CT regime. On the other hand, the EW of the $\rm{FeK\alpha}$ line shows a mean value of $\rm{<log(EW(FeK\alpha))> = 2.89}$ (Fig.\,\ref{fig:nhvseq}, top histogram), with 20 sources presenting upper limits.
Using \textit{Suzaku} and \textit{Swift}/BAT data, \citet{Kawamuro-16} also study a sample of 10 inefficient AGN. Among the seven objects in common with our sample, three (NGC\,3718, NGC\,4941 and NGC\,5643) show different values for $\rm{N_H}$ (discrepancies around 10-20\%). We find higher $\rm{N_H}$ for the three sources, even when accounting for error bars. However, note that although the model is quite similar, they allow the reflection fraction in {\tt pexrav} and the constant (which they relate to the scattering fraction) free to vary in all cases. The main difference is that we test whether letting the covering fraction free to vary significantly improves the fit. This might explain the $\rm{N_H}$ differences found in both analyses. 

 \citet{Marchesi-18} analyze a sample of 26 sources classified as CT, using \textit{NuSTAR} along with other X-ray facilities. Among those objects, we find 14 sources in common with our work, from which 10 have similar value of $\rm{N_H}$ compared to ours, while three have small discrepancies on the $\rm{N_H}$ (i.e. lower than 10\%), and one case (NGC\,424) for which we have strong differences: they report a reflection dominated source, with $\rm{N_H} = 24.4$, while from our analysis it should be considered as highly obscured ($\rm{log(N_H)=23.1}$). Indeed, this source is also reflection dominated according to our f-test analysis. Thus, statistically speaking, the obscured power-law associated with the intrinsic continuum is not required by the data. Moreover, their reflection model is different to ours, which might be responsible for the small changes in the spectral parameters. 

\citet{Panagiotou-19} also studied six sources in common with this work (M\,106, NGC\,1194, NGC\,4941, NGC\,5643, NGC\,5728, and NGC\,7582). All but two (NGC\,4941 (24.09$\pm$0.15) and NGC\,1194 (23.7$\pm$0.1) present consistent $\rm{N_H}$ values. NGC\,4941 and NGC\,1194 present higher $\rm{N_H}$ compared to ours, even when accounting for error bars. Again, they classify these two sources as reflection dominated so this might be affecting the constraint of the LOS absorption due to a weak detection of the instrinsic continuum. In a second paper of the series, \citet{Panagiotou-20} include \emph{NuSTAR} data for other six sources in common with our analysis (NGC\,1052, NGC\,2655, M\,106, NGC\,5252, NGC\,5283 and NGC\,7213), also showing consistent $\rm{N_H}$. Note that \citet{Panagiotou-20} exclude in their sample selection 21 sources in common with our sample; four (M\,81, NGC\,3998, NGC\,7130, and NGC\,7479) are excluded due to their LINER nature, two (IC\,751 and NGC\,7582) due to variability, and the remaining 15 are excluded for being reflection dominated \citep[although included in][]{Panagiotou-19}.

We plot the expected correlation between the $\rm{N_H}$ and the $\rm{EW(FeK\alpha)}$ emission line in the large panel of Fig.\,\ref{fig:nhvseq}. {The equivalent width increases when the $\rm{N_H}$ does so}, with a Pearson's linear relation coefficient of 0.85 (linear fit shown as dotted brown line in the figure). This is well explained by the predicted EW from a uniform shell of material encompassing the continuum source \citep[blue long-dashed line in large panel of Fig.\,\ref{fig:nhvseq}, see also Fig.\,2 in ][]{Leahy93} . {Moreover, \citet{Guainazzi-05} propose a slight change in this correlation when the reflection occurs in the inner walls of an optically-thick matter, as seen through the unobscured LOS, while the continuum is obscured by material along the LOS \citep[see pink dashed line in Fig.\,\ref{fig:nhvseq} and Fig.\,9 in][]{Guainazzi-05}}. Note that the dispersion on this relation is naturally explained through the differences in geometry and composition of the reflector. Therefore, this reinforces the good estimate on the $\rm{N_H}$ in our analysis. 
{This is also reproduced in more recent models with spherical and torus-like geometries \citep[e.g][]{Ikeda-09, Tanimoto-19}  Moreover, we explore how the EW is affected by the amount of obscuration and the strength of the line (see shaded area in Fig.\,\ref{fig:nhvseq}). We choose NGC\,1052 as a test object and vary the normalization of the FeK$\alpha$ line (from $\rm{norm_{6.4 keV}} = 0.5$ to $\rm{norm_{6.4 keV}} = 5$), from left to right, respectively, while changing the $\rm{N_H}$ in order to see how the relation should change. }We find that there is a value of $\rm{EW}$ above which this correlation saturates and thus we would only measure reflection (i.e., in the cases of reflection dominated sources). The main reason for the differences between this work and \citet{Guainazzi-05} is the fact that they do not have very obscured sources in their sample and thus their best fit does not cover such range. Thus, we confirm that 69 out of the 81 objects ($\rm{\sim 85\%}$) are obscured at X-rays (i.e. $\rm{N_{H}>3\times10^{22}cm^{-2}}$). This is further discussed in Section\,\ref{sec:discussion}.

Furthermore, more than half of the sources (42) appear to be completely covered, 27 present a covering factor $\rm{f_{cov} > 0.5}$, while seven show $\rm{f_{cov} < 0.5}$. 
The remaining five AGN present either upper limits or non-constrained estimates for the covering factor. Among these objects, all of them present either low $\rm{N_H}$ or upper limits whereas the sources presenting full covering, also present high obscuration (with only 6 sources being mildly obscured). 

\section{Discussion}\label{sec:discussion}

The aim of this work is to study the {behaviour} of the reflection component and obscuration at X-rays (classically associated with the torus) at low accretion rates. In order to trace such a component, we selected a sample of 81 sources with \textit{NuSTAR} observations among all the nearby AGN observed so far with available $\rm{M_{BH}}$ measurements. We performed a spectral analysis with a model that accounts for the reflection component from neutral material plus a partially absorbed intrinsic continuum.

Early X-ray observations of local AGN have served as observational proofs of the AGN unified theories, showing that most type-1 AGN are unobscured while type-2 AGN tend to show obscuration exceeding $\rm{N_H>10^{22}cm^{-2}}$ \citep{Awaki91,Bassani99}. However, since then, several authors have shown that the fraction of obscured sources might depend on the evolutionary stage of the sources \citep[see][for a review]{Ramos-Almeida17}.

\begin{figure}
    \includegraphics[width = 1\columnwidth]{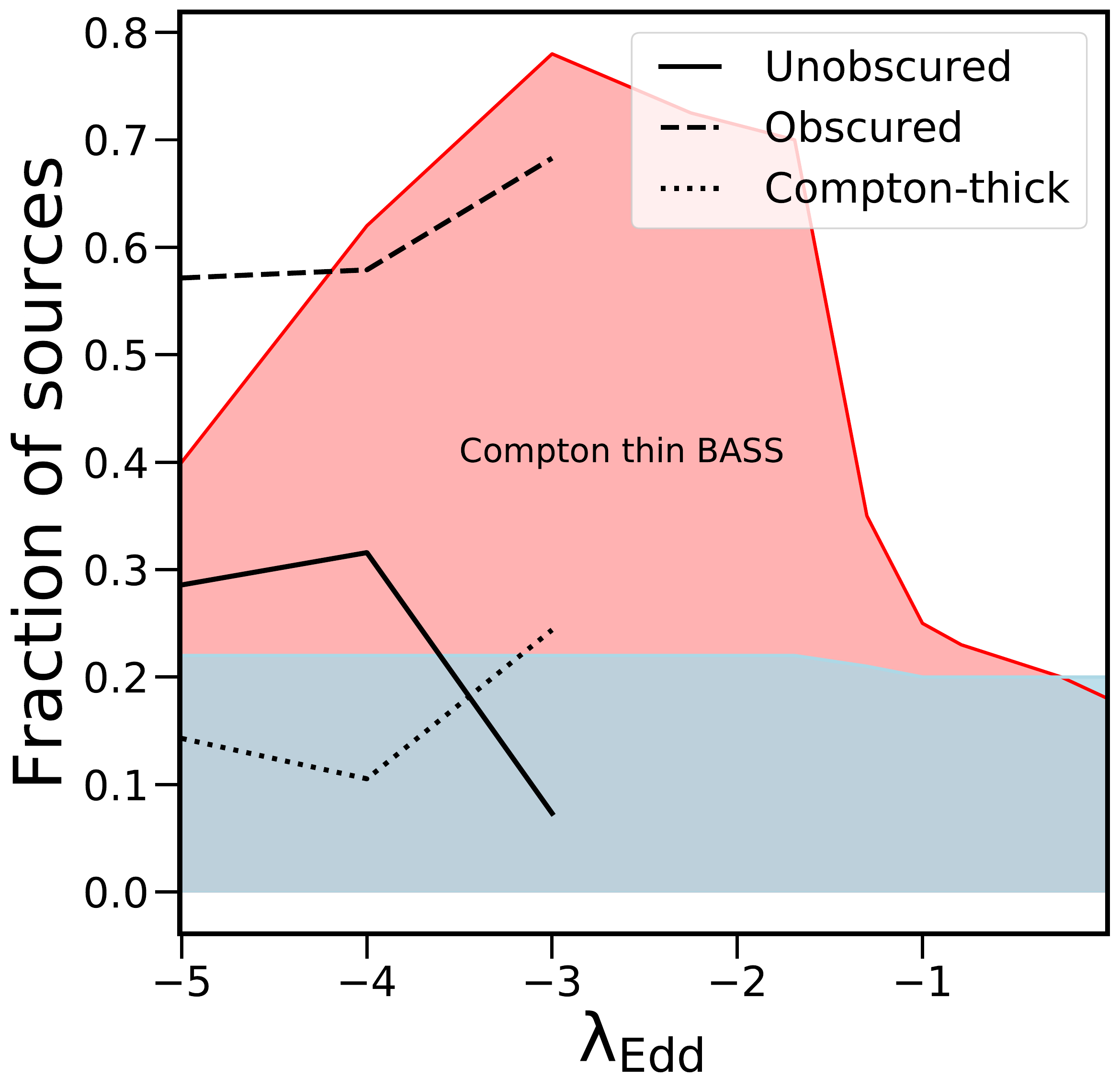}
    \caption{{Fraction of Compton-thin and Compton-thick sources for the \emph{Swift}/BAT sample reported by \citet{Ricci-17} (red and blue shaded areas, respectively) compared to the unobscured, obscured and CT (solid, dashed and dotted lines, respectively). Note that our sample is limited above $\rm{\lambda_{Edd} > -2.5}$ due to the selection criteria we use in this work.}}
    \label{fig:comparison-CTsources}
\end{figure}

The covering factor of the gas can be estimated studying the absorption properties of large samples of AGN. In Fig.\,\ref{fig:comparison-CTsources} we compare the results found by \citet{Ricci-17} using the BASS sample and those found in this study (numbers are reported in Table\,\ref{tab:comparison}). Using this technique, recent hard X-ray studies have shown that the fraction of CT sources below $\rm{\lambda_{Edd} \sim0.032}$ is $\rm{ 23\pm 6 \%}$ (blue shaded area in Fig.\,\ref{fig:comparison-CTsources}). \citet{Marchesi-18} find that the fraction of CT sources decreases when using \emph{NuSTAR} data thanks to a better covering and sensitivity above 10\,keV which allows a better restriction in the LOS obscuration affecting the intrinsic continuum.

 Our results indicate that the fraction of CT sources at $\rm{\lambda_{Edd}\simeq 10^{-5}}$ is $\rm{15\%}$ (dotted line), while \citet{Ricci-17} find it to be $\rm{\sim 20\pm 4\%}$. Thus, our results are statistically in agreement with them. Meanwhile, \citet{Ricci-17} find that the peak of obscured sources happens around $\rm{\lambda_{Edd} \sim 10^{-3}}$, rising from 20\% to 80\% at Eddington rates $\rm{\lambda_{Edd}\simeq 10^{-1}}$ and $\rm{\lambda_{Edd}\simeq 10^{-3}}$, respectively \citep[see red shaded area in Fig.\,\ref{fig:comparison-CTsources}, see also][]{Ricci-17}. Interestingly, our sample agrees well with these numbers at $\rm{\lambda_{Edd}\simeq 10^{-3}}$ for the percentage of obscured AGN (dashed line in Fig.\,\ref{fig:comparison-CTsources}). Note that our results indicate a low population of unobscured sources at $\rm{\lambda_{Edd}\simeq 10^{-3}}$. Moreover, a slight decrease on the fraction of Compton-thin obscured sources is found below $\rm{\lambda_{Edd}\simeq 10^{-3}}$, reaching $\sim$60\% at $\rm{\lambda_{Edd}\simeq 10^{-5}}$. Correspondingly, the fraction of unobscured sources rises from $\rm{\sim 6}$ at $\rm{\lambda_{Edd}\simeq 10^{-3}}$ to $\rm{\sim 30}$\% at $\rm{\lambda_{Edd}\simeq 10^{-5}}$. Nonetheless, we are aware that this may be due to a bias in the selection criteria. Indeed, if there were more unobscured sources at these Eddington rates, they should have been included in our sample. However, we select only those sources with archival data and a minimum S/N in the spectrum. This has two main consequences: i) there may be several objects which have not yet been observed, among which there might be unobscured sources. ii) there might be more obscured sources for which the flux is so low that they do not fulfill our S/N criteria. Thus, although our results are in agreement with the complete sample by \citet{Ricci-17}, we warn the reader that our sample is incomplete.

The trend found in this analysis might suggest no obscuration for a large fraction of sources at even lower accretion rates. The scenario in which both the BLR and the torus can be seen as different regions from a wind coming off the disk has long been proposed \citep{Elvis-00} and it naturally includes the absence of such regions in some objects \citep[e.g., True-type 2 AGN intrinsically miss the BLR][]{Tran01,Laor-03,Panessa-06}. If the accretion disk cannot expel material in an efficient way, then the gravitational force from the SMBH will overcome the radiation pressure from the wind, causing the wind, and therefore the BLR and the torus, to collapse \citep{Elitzur-06}. Therefore, as a first approach, the condition for the torus disappearance can be related to the bolometric luminosity of the source and the accretion rate. According to \citet{Elitzur-06}, the BLR should disappear for low-luminosity AGN (i.e., $\rm{L_{bol} < 10^{42} \ erg \ s^{-1}}$). Six out of the 11 objects lacking reflection signatures in our sample are below this threshold. However, in a more recent study it has been shown that this structure can disappear even for bright sources \citep[e.g.][]{Elitzur-16}, depending on the properties of the wind, such as the wind efficiency or the number/density of clouds. In fact, it requires a minimal column density, implying a minimal outflow rate and thus a minimal accretion rate. Interestingly, three of our candidates present upper limits on the $\rm{N_H}$ and other five show $\rm{N_H \leq 10^{23}cm^{-2}}$. Finally, three are rather obscured (namely M\,106, NGC\,612 and M\,88).
 In particular, for M\,106, \cite{Kawamuro-16} find that the $\rm{FeK\alpha}$ line is variable, proposing it might be originated in the accretion disk, while in the case of NGC\,612, \citet{Ursini-18} do not find strong features of a Compton-reflection component, although like us, they find that this source presents high absorption, proposing that it might originate in an extended structure rather than the pc-scale torus. 
Thus, these works are still in agreement with the lack of reflection originated in the torus. Another particularly interesting object is M\,88, being proposed as a True-Type 2 AGN \citep[i.e., intrinsically missing the BLR,][]{Brightman-08}. We would expect for this object to be low-obscured. Although we modelled this source with an obscured component, the quality of the observations\footnote{The spectra of M\,88 show calibration issues clearly visible in Fig.\,\ref{fig:candidates} and resulting in a calibration factor much larger than for the rest of the sample, with a value of $\rm{C_{cross} = 1.99\pm0.19}$.} may prevent us from finding a reliable value in the $\rm{N_H}$ for this particular source. Thus, almost all our candidates to the torus disappearance are on the lower end of LOS obscuration (Fig.\,\ref{fig:nhvseq}).

\begin{figure}
\centering
    \includegraphics[width=1\columnwidth]{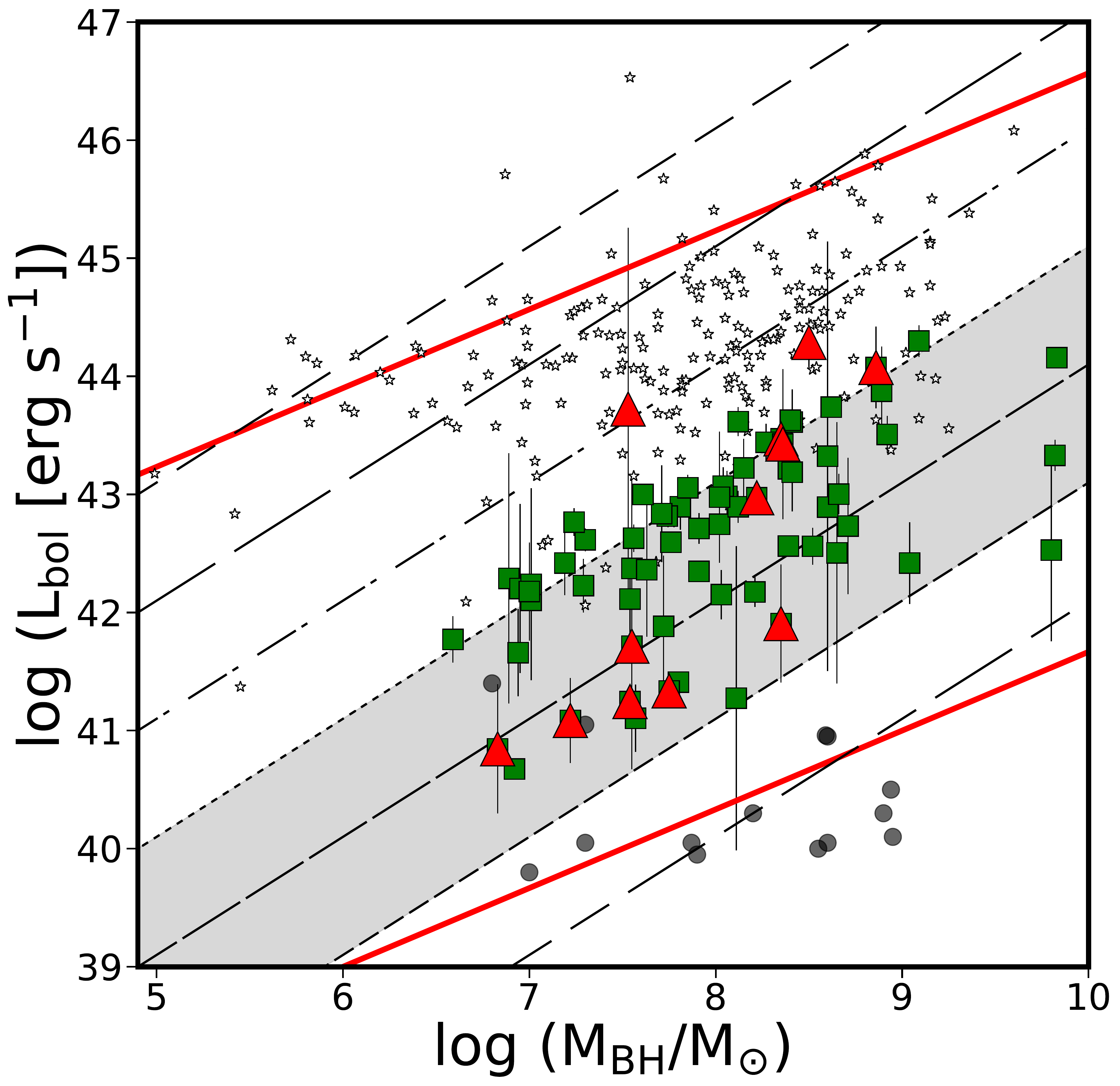}
   
    \caption{Bolometric luminosity versus $\rm{M_{BH}}$ for our sample (colors and symbols as in Fig.\,\ref{fig:reflvslfe}) and the BASS sample (small empty stars). Our 11 candidates are plotted in the larger red triangles. The semi-translucent circles represent the candidates for torus disappearance found by \citet{Gonzalez-17} using mid-IR data. { The black dashed lines represent $\rm{\log(L_{Edd}) = [0, -1, -2, -3, -4, -5, -6]}$ from top to bottom, respectively}. The two red lines enclose the region in which the torus may or may not exist depending on the wind parameters, as proposed by \citet{Elitzur-09}. The gray-shaded area corresponds to the Eddington rates from the sample not biased by the selection criteria. Note that those sources with Eddington rates above $\rm{\log \ \lambda_{Edd} \sim -2.5}$ are not shown in this plot since at those Eddington rates our sample is not complete.  }
    
     \label{fig:LbolvsMbh}
  \end{figure}

According to \citet{Elitzur-16}, the BLR/torus may or may not exist in a delimited zone in the luminosity-$\rm{M_{BH}}$ diagram and both components should disappear under a certain threshold delimited by the wind parameters. Fig.\,\ref{fig:LbolvsMbh} shows the bolometric luminosity versus $\rm{M_{BH}}$ diagram for our sample (large symbols), excluding the sources for which $\rm{\lambda_{Edd} > -2.5}$  Note that although these sources are mostly CT at this Eddington rate, it is difficult to asses how many sources we may be missing due to the selection biases we have imposed and to the available data.

As a comparison sample, we also plot data from the \textit{Swift}/BAT sample \citep[][plotted as the small star symbols]{Koss-17}. 
We selected sources from that sample such that they had some estimate on the ${\rm{M_{BH}}}$, with reported values for the intrinsic 2-10\,keV luminosity, to obtain the bolometric luminosity and LOS obscuration. We discarded sources without constrained  $\rm{\log(N_H)}$ values and we also neglect the sources in common. We obtained a total of 195 sources for the comparison sample. Note that all our sources fall within the region in which the BLR/torus may or may not exist (red lines), according to \citet{Elitzur-16}. We also plot with red triangles our 11 candidates for the torus disappearance. The location of our candidates is in complete agreement with the idea that the torus disappearance does not rely solely on the bolometric luminosity of the source, yet, it does disappear for inefficient sources. 
Moreover, \cite{Gonzalez-17} study a sample of AGN using mid-IR spectra trying to understand the behaviour of the torus in what is called the gradual resizing. We plot in Fig.\,\ref{fig:LbolvsMbh} (gray circles) their candidates for the disappearance of the torus from a mid-IR point of view. Unlike us, they do have objects lacking the torus below the limit proposed by \citet{Elitzur-16}, at $\rm{\lambda_{Edd}=10^{-6}-10^{-5}}$ depending on the $\rm{M_{BH}}$. However, they also found several candidates in the range between $\rm{\lambda_{Edd}=10^{-5}-10^{-3}}$. Our study complements theirs by the inclusion of these 11 new candidates to the torus disappearance; all together a sample of 25 AGN. Indeed, we may also be missing objects with $\rm{\lambda_{Edd} < 10^{-5.5}}$, either because observations for such low-luminosity objects have not yet been performed, or because the S/N of the observations are so low that they do not fulfill our selection criteria. Moreover, highly obscured objects might be missing because they are more difficult to detect than unobscured objects.
\begin{table}
    \centering
    \begin{tabular}{ccccc}
 
\hline \hline
    $\rm{log(\lambda_{Edd})}$ (total) & Unobs. & Obs & CT & No ref.\\
     & \# (\%) & \# (\%) & \# (\%) & \# (\%)  \\
     (1) & (2) & (3) (4) & (5)\\
    \hline 
-5 (7) & 2 (28)& 4 (57) & 1 (14) & 2 (28)   \\
-4 (19) & 6 (31) & 11 (58) & 2 (10) & 4 (21)  \\
-3 (41) & 3 (7) & 28 (68) & 10 (24) & 3 (7) \\
\hline        
    \end{tabular}
    \caption{ Number of unobscured, obscured, CT and no reflection sources per Eddington rate range. In parenthesis it is represented the contribution percentage for each of the classifications. Note that sources with no reflection do not sum up to the total contribution in each range, since both the $\rm{N_H}$ classification is independent.} 
    \label{tab:comparison}
\end{table}

Another interesting result on the obscurer {behaviour} is the increase on the narrow FeK$\rm{\alpha}$ line luminosity compared to the reflection luminosity {when} the reflection luminosity decreases (Fig.\,\ref{fig:reflvslfe}, large right panel). This is present when we compare the reflection luminosity with the intrinsic luminosity, showing a {mean contribution of 40\% of reflection} for the full range of the intrinsic luminosity (Fig.\,\ref{fig:reflvslfe}, large left panel). However, {the FeK${\rm{\alpha}}$ line and reflection component changes} are not associated with the accretion rate, because no correlation is found between the accretion rate and the luminosity of the reflection component or the FeK$\rm{\alpha}$ line. Although the flux of the narrow FeK$\rm{\alpha}$ line (compared to the intrinsic X-ray flux) is generally weaker in type-2 AGN than in type-1 AGN for the same 10-50 keV continuum luminosity \citep{Ricci14}, this is not reproduced by the low accretion AGN in our sample. \citet{Ricci14} suggest that this difference can be explained by means of different average inclination angles with respect to the torus, as predicted by the unified model. However, our sample is mostly constituted by type-2 AGN which might explain why we do not find such behaviour. While the FeK$\rm{\alpha}$ line can be produced by material with column densities as low as $\rm{N_H \simeq 10^{21-23}\,cm^{-2}}$, the Compton hump can only be created by the reprocessing of X-ray photons in CT material. Thus, the increase on the fraction of FeK$\rm{\alpha}$ line luminosity compared to the Compton hump luminosity toward low reflection luminosities might be an indication of a higher fraction of the emission produced in Compton-thin material associated with the accretion disk. If the accretion disk contribution is responsible for this behaviour, we might expect differences on the photon index distributions that might contribute to the Compton hump and or the FeK$\rm{\alpha}$ line differently as long as the reflection luminosity decreases. For instance, it is expected for the photon index of standard disks to be larger for higher intrinsic luminosities (softer-when-brighter), while it is expected to decrease with increasing luminosity in radiatively inefficient accretion disks \citep[harder-when-brighter,][and references therein]{Narayan-95, Connolly-16}. 

However, the mean of the distribution of the photon index ($\rm{< \Gamma > = 1.81}$) is similar to the average photon index found in diverse AGN samples \citep[e.g. \emph{Swift}/BAT,][]{Koss-17}, and we do not find any {particular trend} on the accretion disk that might explain this behaviour found in the strength of the line. Another alternative is a smooth transition on the average iron abundances of the X-ray reflector \citep[i.e. torus,][]{Ricci14}. This might imply a smooth transition on the chemistry of the torus as seen at X-rays as long as the torus luminosity decreases traced by the Compton hump. In favor, \cite{Gonzalez-17} also find that a group of objects with a low contribution from the torus, are not well described with the clumpy torus model at mid-IR. They suggest that this might imply that the torus chemistry may be different at low luminosities, consistent with these findings.

\section{Summary and conclusions}
\label{sec:conclusions}

We have performed an analysis aimed to find the conditions for the {change} and plausible disappearance of the obscuring gas for low-accretion AGN. For that purpose, we used a sample of 81 sources observed with \textit{NuSTAR} selected by having Eddington rates $\rm{\lambda_{Edd}\lesssim 10^{-3}}$. We fitted the data with a model accounting for a patchy absorber and a reflection component from neutral material. We found that $\sim80\%$ of our sample is obscured, with $20\%$ of CT sources, in agreement with several works \citep[e.g.][]{Ricci-17, Marchesi-18}. 

In order to directly compare the lack of reflection component with those sources lacking of the torus at mid-IR wavelengths, fainter X-ray sources, with luminosities $\rm{L_{X} \sim 10^{38} \ erg \ s^{-1}}$ are required. {However}, we found that a small number of sources in our sample ($\sim$15\%) seem to lack the reflection component associated with the torus, while the rest of the sources are found to have reflection features such as the existence of the $\rm{FeK\alpha}$ line and/or the reflection hump. The sources lacking of reflection component features are located in the region where the torus may or may not disappear according to their $\rm{M_{BH}}$ and bolometric luminosities. When the reflection component was detected, it accounts for $40\%$ of the luminosity of the intrinsic continuum. We also found tentative evidence in favour of an increase of unobscured sources and a change of the reflector chemistry toward $\rm{\lambda_{Edd}\sim 10^{-5}}$. Our findings are in agreement with the scenario in which as the AGN becomes less efficient, the torus is not supported by the radiation field from the wind coming off the accretion disk and fades. As a final remark, we highlight the importance of better S/N data and a {broader energy} coverage in order to apply more complex models for the reflection component, which includes testing if it may also be produced in the accretion disk (i.e., ionized reflection), but also testing different geometries and matter distributions (i.e, homogeneous, clumpy, etc).

\section*{Acknowledgements}

We thank the anonymous referee for her/his useful comments which greatly improved this paper. This research has made use of the NASA/IPAC Extragalactic Database (NED), which is operated by the Jet Propulsion Laboratory, California Institute of Technology, under contract with the National Aeronautics and Space Administration. This research has made use of data and/or software provided by the High Energy Astrophysics Science Archive Research Center (HEASARC), which is a service of the Astrophysics Science Division at NASA/GSFC and the High Energy Astrophysics Division of the Smithsonian Astrophysical Observatory. NOC would like to thank CONACyT scholarship No. 897887. CVC acknowledges support from CONACyT. We thank the UNAM PAPIIT project IN105720 (PI OGM). LHG acknowledges funds by ANID – Millennium Science Initiative Program – ICN12$\_$009 awarded to the Millennium Institute of Astrophysics (MAS).

\section*{Data Availability}

The data underlying this article are available in at \url{https://heasarc.gsfc.nasa.gov/}. The datasets were derived from sources in the public domain.


\appendix

\section{Modeling with CABS}
\label{Appendix-A}

\begin{figure}
    \centering
    \includegraphics[width = 1\columnwidth]{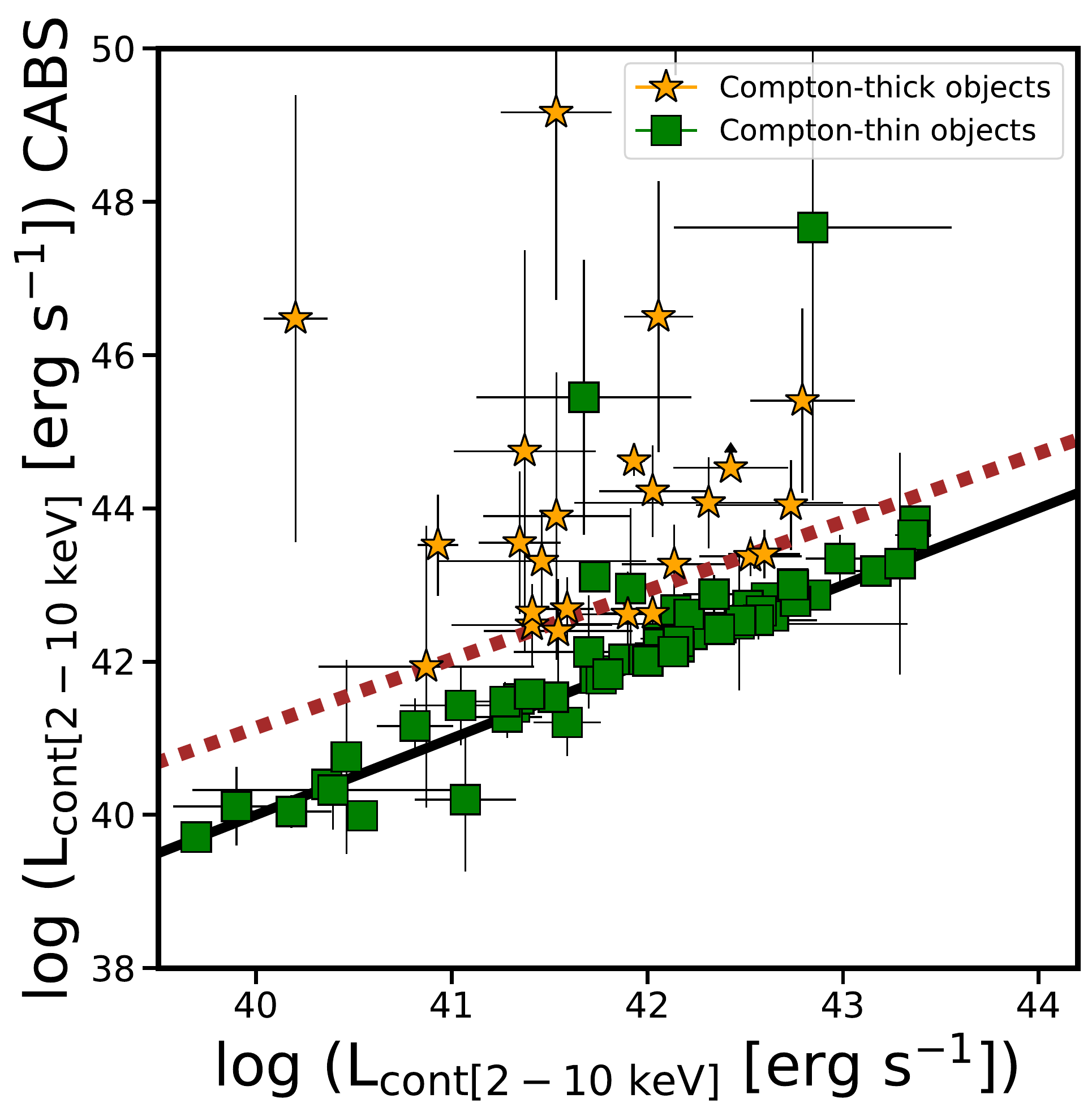}
    \caption{Intrinsic luminosity of the power-law component when including {\tt cabs} in the spectral modeling (vertical axis) compared to the case in which it is not used (horizontal axis). The brown dotted and black solid lines are the best fit to the data and the 1:1 relation, respectively. }
    \label{fig:cabs-lum}
\end{figure}

We use a model which does not account for the Compton scattering, particularly important for Compton-thick objects \citep{Annuar-17, Ricci-17, Zhao-19, Marchesi-19}. {\sc xspec} has a model to account for Compton scattering called {\tt cabs}\footnote{see {\sc xspec} manual \url{https://heasarc.gsfc.nasa.gov/xanadu/xspec/manual/XspecManual.html}} . This model component has been reported to present some issues  \citep[see e.g. {\sc mytorus} manual and][]{Tanimoto-18}{, and does not emulate properly the X-ray spectra of AGN.} 
However, {\tt cabs} has been used in several recent works to account for Compton scattering in AGN when using \emph{NuSTAR} data \citep[see ][]{Ricci-17, Oda-17, Tanimoto-20}. In practice, we added the {\tt cabs} component to the model in Eq. (4) as follows:
\begin{multline}
M_{5} =  {\tt phabs_{\rm{Gal}}}
(({\tt zphabs_{\rm{intr}}}*{\tt cabs}*{\tt zpowerlw}) +  {\tt ct}*{\tt zpowerlw} \\
+ {\tt pexmon} + {\tt zgauss_{6.7 \ keV}} + {\tt zgauss_{6.97 \ keV}})
\end{multline}

\noindent letting the $\rm{N_H}$ in the {\tt cabs} component to be the same as that of the LOS obscuration. The resulting $\rm{N_H}$ values are mostly consistent with those reported in this work (see below). However, the intrinsic luminosity of the sources obtained for some objects changes  by a factor of $10^4$ (see Fig.\,\ref{fig:cabs-lum} shows). Indeed, note that a non-negligible amount of sources fall well outside the 1:1 relation even when accounting for error bars (black solid line). Moreover, the error bars are much larger when using {\sc cabs} than when excluding it. These luminosities result in Eddington rates close to the Eddington limit, which are unrealistic values for LLAGN.

{These results imply that there is a systematic uncertainty when adding the Compton-scattering through the {\tt cabs} component. However, this does not imply that such a component does not physically exist in the X-ray spectra of AGN. In fact, the neglection of this component does add a systematic uncertainty to the general analysis. Indeed, this effect primarily affects the X-ray spectra of CT objects. Thus, although the addition of {\tt cabs} does not lead to physically realistic results, the exclusion of the Compton-scattering effects in our analysis, imply that all intrinsic luminosities for CT sources reported here should be treated as lower limits. }

\begin{figure}
    \centering
    \includegraphics[width = 1\columnwidth]{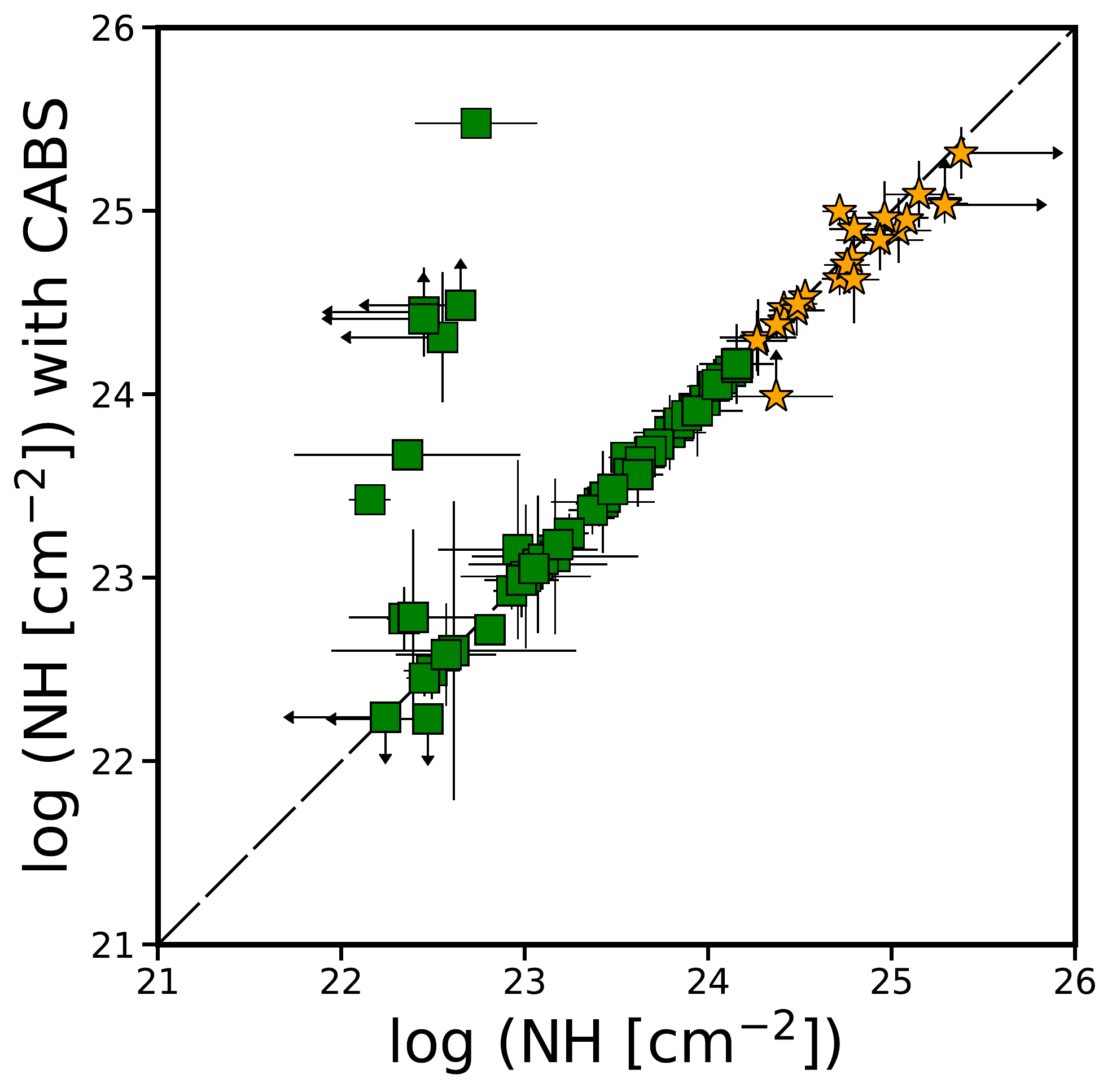}
    \caption{Column density of the sources when {\tt cabs} is added (vertical axis) compared to the case in which it is not used (horizontal axis). The black dashed line is the 1:1 relation.}
    \label{fig:nhcabs}
\end{figure}

We have also checked how the exclusion of this component affects the $\rm{N_H}$ measurement. Fig.\,\ref{fig:nhcabs} shows the change in this parameter. The black dashed line shows the 1:1 relation. In this case, most of the sources fall in this line, with only 8 sources falling outside the relation. Note that this is a result similar to the one by \citet[][see their Fig. 5a]{Tanimoto-18} , where they found larger values for the intrinsic luminosity when using this model. However, the spectral parameters were not very much affected. 

\label{lastpage}

\bsp	

\end{document}